\documentclass[11pt]{article}
\usepackage[utf8]{inputenc}
\usepackage[english]{babel}
\usepackage{multicol}
\usepackage{epsfig,amssymb,amsmath, euscript,color, mathrsfs, graphicx}
\renewcommand{\baselinestretch}{1.3}
\makeatletter

 \@addtoreset{equation}{section}
 \renewcommand{\theequation}{\thesection.\arabic{equation}}
\newtheorem{theorem}{Theorem}
\newtheorem{lemma}{Lemma} 
\newtheorem{corollary}{Corollary}

\renewcommand{\hat}{\widehat}
\def\singlespace{\def\baselinestretch{1}\@normalsize}

\def\wh{\widehat}
\def\wt{\widetilde}
\def\askip{\vspace{0.1in}}

\newcommand{\MAE}{{\rm MAE}}

\newcommand{\cov}{{\rm Cov}}
\newcommand{\diag}{{\rm diag}}

\newcommand{\var}{\mbox{Var}}

\def\la{\lambda}

\newcommand{\bA}{{\mathbf A}}
\newcommand{\bB}{{\mathbf B}}

\newcommand{\bH}{{\mathbf H}}
\newcommand{\bI}{{\mathbf I}}

\newcommand{\bS}{{\mathbf S}}
\newcommand{\bU}{{\mathbf U}}
\newcommand{\bV}{{\mathbf V}}
\newcommand{\bW}{{\mathbf W}}
\newcommand{\bX}{{\mathbf X}}
\newcommand{\bY}{{\mathbf Y}}
\newcommand{\bZ}{{\mathbf Z}}
\newcommand{\ba}{{\mathbf a}}
\newcommand{\bb}{{\mathbf b}}

\newcommand{\be}{{\mathbf e}}

\newcommand{\bv}{{\mathbf v}}
\newcommand{\bw}{{\mathbf w}}
\newcommand{\bx}{{\mathbf x}}
\newcommand{\by}{{\mathbf y}}
\newcommand{\bz}{{\mathbf z}}

\newcommand{\blambda}{\boldsymbol{\lambda}}

\newcommand{\bSigma}{\boldsymbol{\Sigma}}

\newcommand{\bve}{\mbox{\boldmath$\varepsilon$}}

\def\6bullets{\bullet\bullet\bullet\bullet\bullet\bullet}

\newcommand{\qqed}{\hfill $\Box$ \vspace{1mm}}

\begin{document}

\title{Generalized Yule-Walker Estimation for Spatio-Temporal\\ Models
with Unknown Diagonal Coefficients}

\author{
Baojun Dou$^{\dagger}$ \quad  Maria Lucia Parrella$^{\ddagger}$  \quad
Qiwei Yao\thanks{Corresponding author. Department of Statistics, London School of Economics, Houghton Street, London, WC2A 2AE, United Kingdom. Tel. :  +44 (0)20 7955 6767. E-mail address: q.yao@lse.ac.uk.}\\
  $^{\dagger,*}$Department of Statistics, London School of
Economics, London, U.K.\\
 $^\ddagger$Department of Economics and Statistics,
University of Salerno,  Fisciano, Italy\\
$^*$Guanghua School of Management, Peking University, Beijing, China
}

\date{}
\maketitle

\begin{abstract}
We consider a class of spatio-temporal models which extend popular
econometric spatial autoregressive panel data models by allowing the
scalar coefficients for each location (or panel) different from each
other. To overcome the innate endogeneity, we propose a generalized
Yule-Walker estimation method which applies the least squares estimation to
a Yule-Walker equation. The asymptotic theory is developed under
the setting that both the sample size and the number of locations (or
panels) tend to infinity under a general setting for stationary and
$\alpha$-mixing processes, which includes spatial autoregressive panel data models
driven by $i.i.d.$ innovations as special cases.
The proposed methods are illustrated using both
simulated and real data.
\end{abstract}

\bigskip\bigskip




\newpage

\section{Introduction}

The class of spatial autoregressive (SAR) models is introduced to
model cross sectional dependence of different economic individuals
at different locations (Cliff and Ord, 1973).
More recent developments extend SAR models to spatial dynamic panel data (SDPD) models,
i.e. adding time lagged terms to account for serial
correlations across different locations. See, e.g. Lee and Yu (2010a).
Baltagi et al. (2003) considers a static spatial panel model where the
error term is a SAR model. Lin and Lee (2010) shows that in the presence
of heteroskedastic disturbances, the maximum likelihood estimator for the
SAR models without taking into account the heteroskedasticity is
generally inconsistent and proposes an alternative GMM estimation
method. Computationally the GMM methods are more
efficient than the QML estimation (Lee, 2001).
Lee and Yu (2010a) classifies SDPD models
into three categories: stable, spatial cointegration and explosive cases.
As pointed out by Bai and
Shi (2011), the cases with a large number of cross sectional units and a
long history are rare. Hence it is pertinent to consider the setting with short time
spans in order to include as many locations as possible. Both estimation
method and asymptotic analysis need to be adapted under this new setting.
Yu et al. (2008) and Yu et al. (2012)  investigate the asymptotic properties
when both the number of locations and the length of time series tend to infinity for both
the stable case and spatial cointegration case, and show that QMLE is 
consistent.

Motivated by the evidence in some practical
examples, we extend the model in Yu et al. (2008) and Yu et al. (2012) by
allowing the scalar coefficients for each location (or panel) different
from each other. This increase in model  capacity comes with
the cost of estimating substantially more parameters.
In fact that the number of the parameters in this new setting is
in the order of the number of locations.
The model considered in this paper has
four additive
components: a pure spatial effect, a pure dynamic effect, a time-lagged spatial
effect and a white noise.
Due to the innate endogeneity, the conventional regression estimation methods such as the least
squares method directly based on the model lead to inconsistent estimators.
To overcome the difficulties caused by the endogeneity,
 we propose a generalized Yule-Walker type estimator for estimating the
parameters in the model, which applies the least squares estimation to
a Yule-Walker equation. The asymptotic normality of the
proposed estimators is established under the
setting that both the sample size $n$ and the number of locations (or
panels) $p$ tend to infinity. Therefore the number of parameters to be
estimated also diverges to infinity, which is a marked difference from,
e.g., Yu et al. (2012).
We develop the asymptotic properties under a general setting for stationary
and $\alpha$-mixing processes, which includes the spatial autoregressive panel data
models driven by $i.i.d.$ innovations as special cases.

The rest of the paper is organized as follows. Section 2 introduces the
new model, its motivation and the generalized Yule-Walker estimation
method. The asymptotic theory for the proposed estimation method is
presented in Section 3. Simulation results and
real data analysis are reported, respectively, in Section 4 and 5. All
the technical proofs are relegated to an Appendix.

\section{Model and Estimation Method}
\subsection{Models}
The model considered in this paper is of the following form:
\begin{equation}\label{fullmodel}
\by_t =D(\blambda_0)\bW\by_t +D({\blambda_1})\by_{t-1}+ D(\blambda_2)\bW\by_{t-1} +\bve_t, \tag{1}
\end{equation}
where $\by_t=(y_{1,t},\ldots,y_{p,t})^{T}$ represents the observations from
$p$ locations at time $t$, $D(\blambda_k)=\diag(\lambda_{k1}, \ldots,
\lambda_{kp})$ and $\lambda_{kj}$ is the unknown coefficient parameter for the
$j$-th location, and $\bW$ is the $p \times p$ spatial
weight matrix which measures the dependence among different locations.
All the main diagonal elements of $\bW$ are
zero.
It is a common practice in
spatial econometrics to assume $\bW$ known. For example, we may let
$w_{ij} = 1 /(1+d_{ij})$, for $i \neq j$, where $d_{ij}\ge 0$ is an
appropriate distance between the $i$-th and the $j$-th location. It can
simply be the geographical distance between the two locations or the
distance reflecting the correlation or association between the variables
at the two locations. In the above model,
$D(\blambda_0)$ captures the pure spatial effect, $D(\blambda_1)$
captures the pure dynamic effect, and $D(\blambda_2)$ captures the time-lagged
spatial effect.
We also assume that the error term
 $\bve_t=(\varepsilon_{1,t}, \varepsilon_{2,t},
\ldots, \varepsilon_{p,t})^T$ in (\ref{fullmodel}) satisfies the condition
$
\mathrm{Cov}\left(\by_{t-1}, \bve_t\right)=0.
$
When  $\lambda_{k1} = \cdots = \lambda_{kp}$ for $k=0, 1, 2$,
(\ref{fullmodel}) reduces to the model of Yu et al. (2008), in which
there are only 3 unknown regressive coefficient parameters.
In general the regression function in (\ref{fullmodel}) contains
$3p$ unknown parameters.

\begin{figure}
        \centering
        \mbox{ \resizebox{3.6cm}{!}{\includegraphics{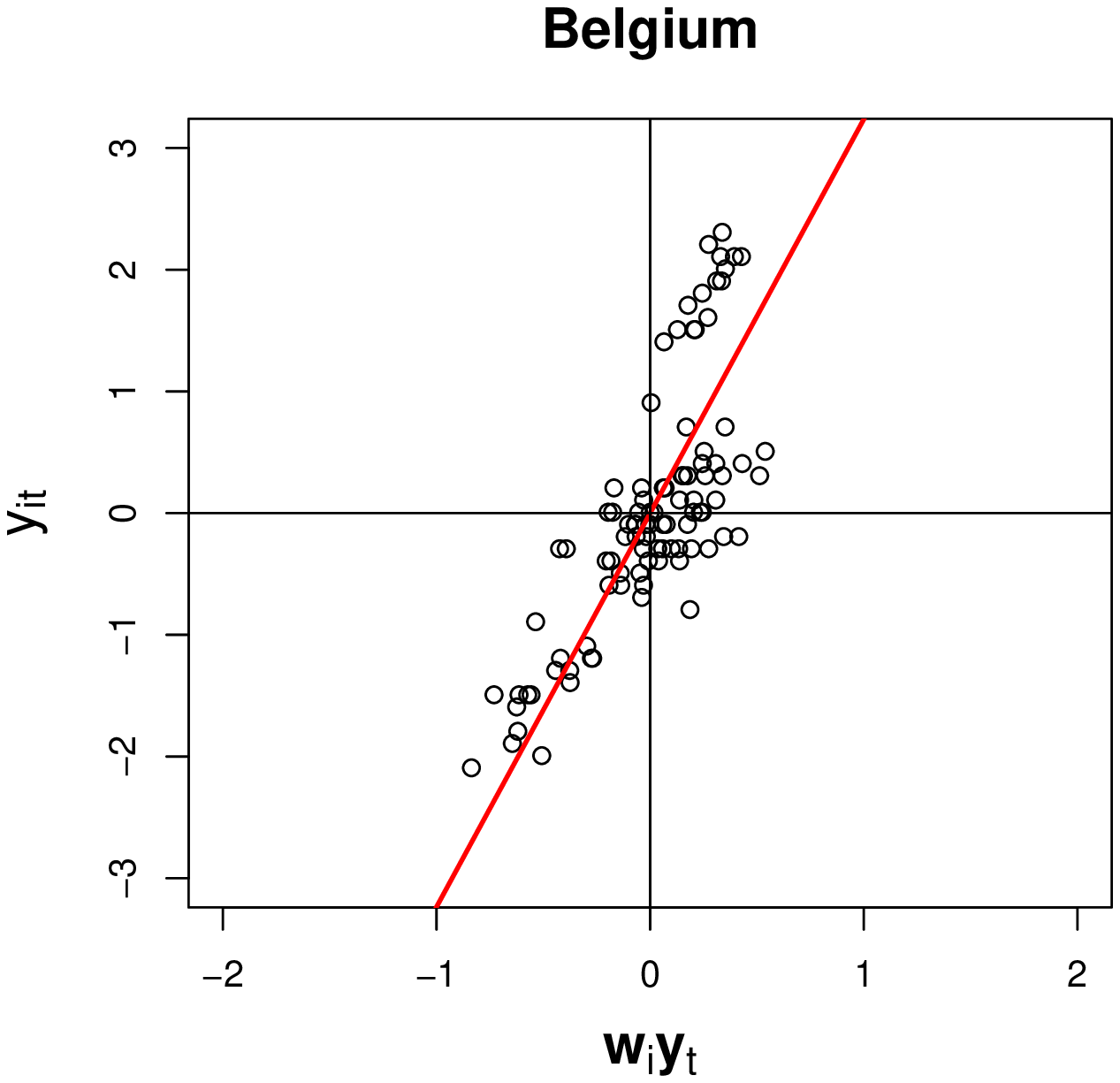}} }
        \mbox{ \resizebox{3.6cm}{!}{\includegraphics{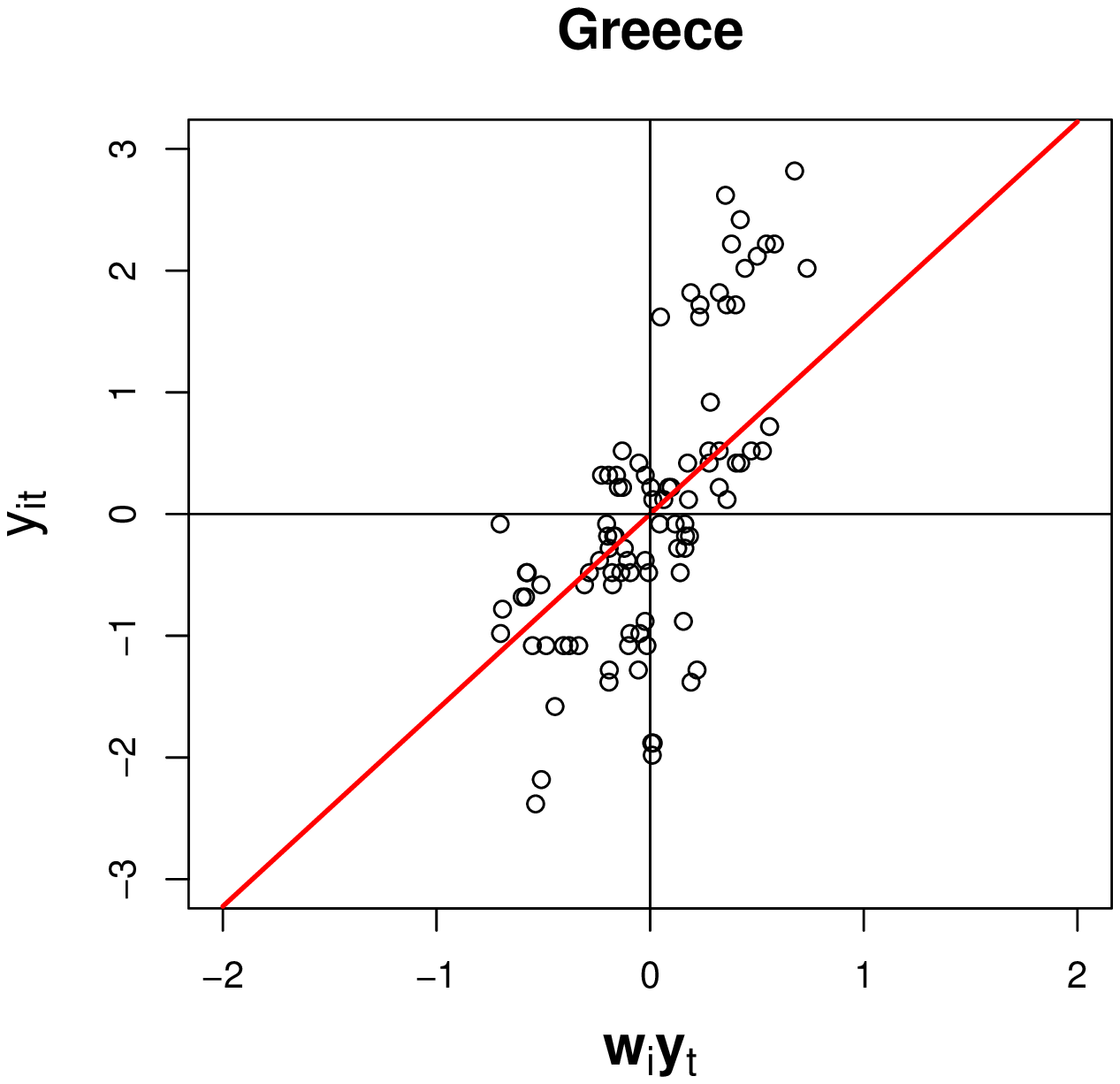}} }
        \mbox{ \resizebox{3.6cm}{!}{\includegraphics{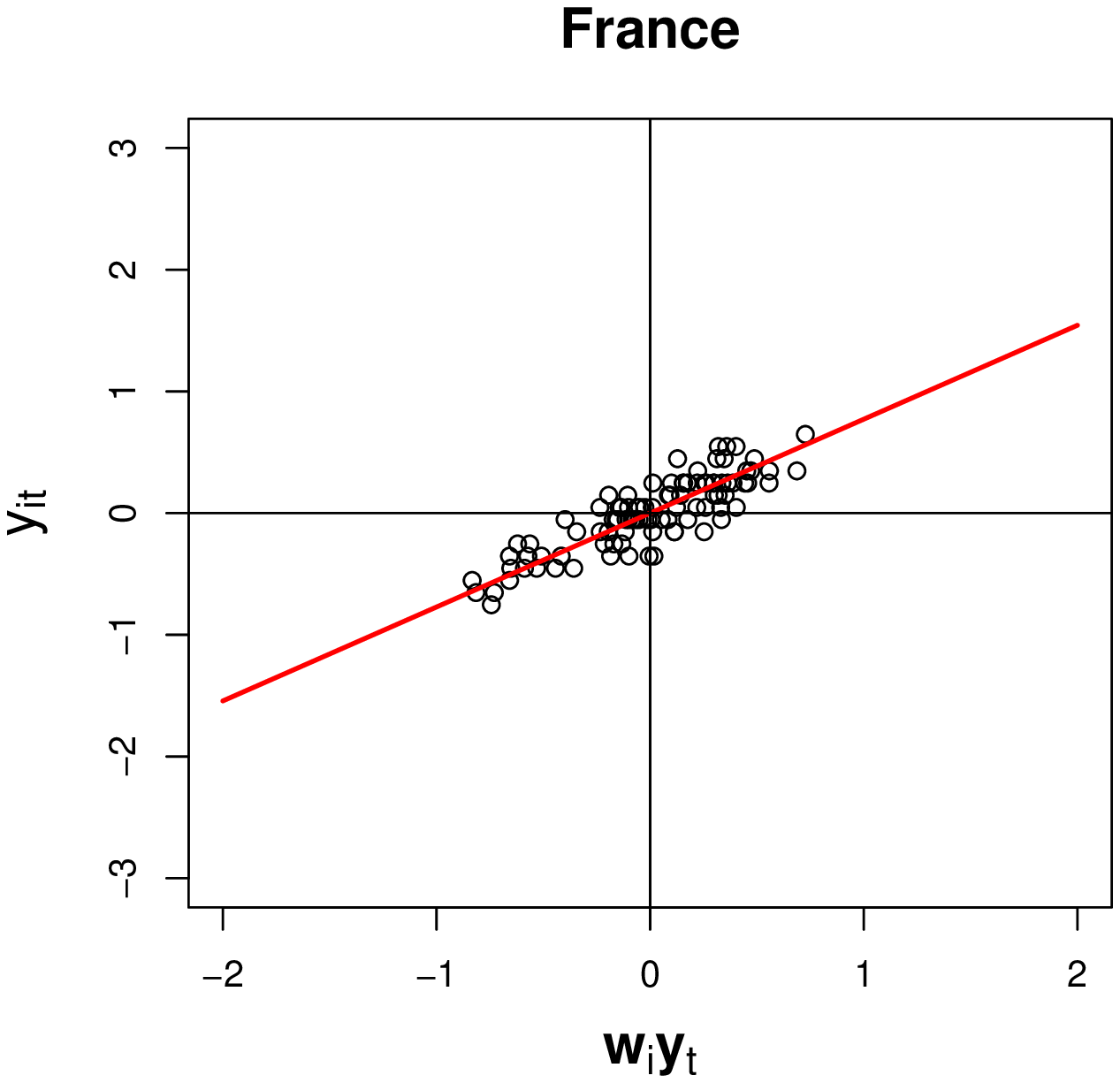}} }
        \mbox{ \resizebox{3.6cm}{!}{\includegraphics{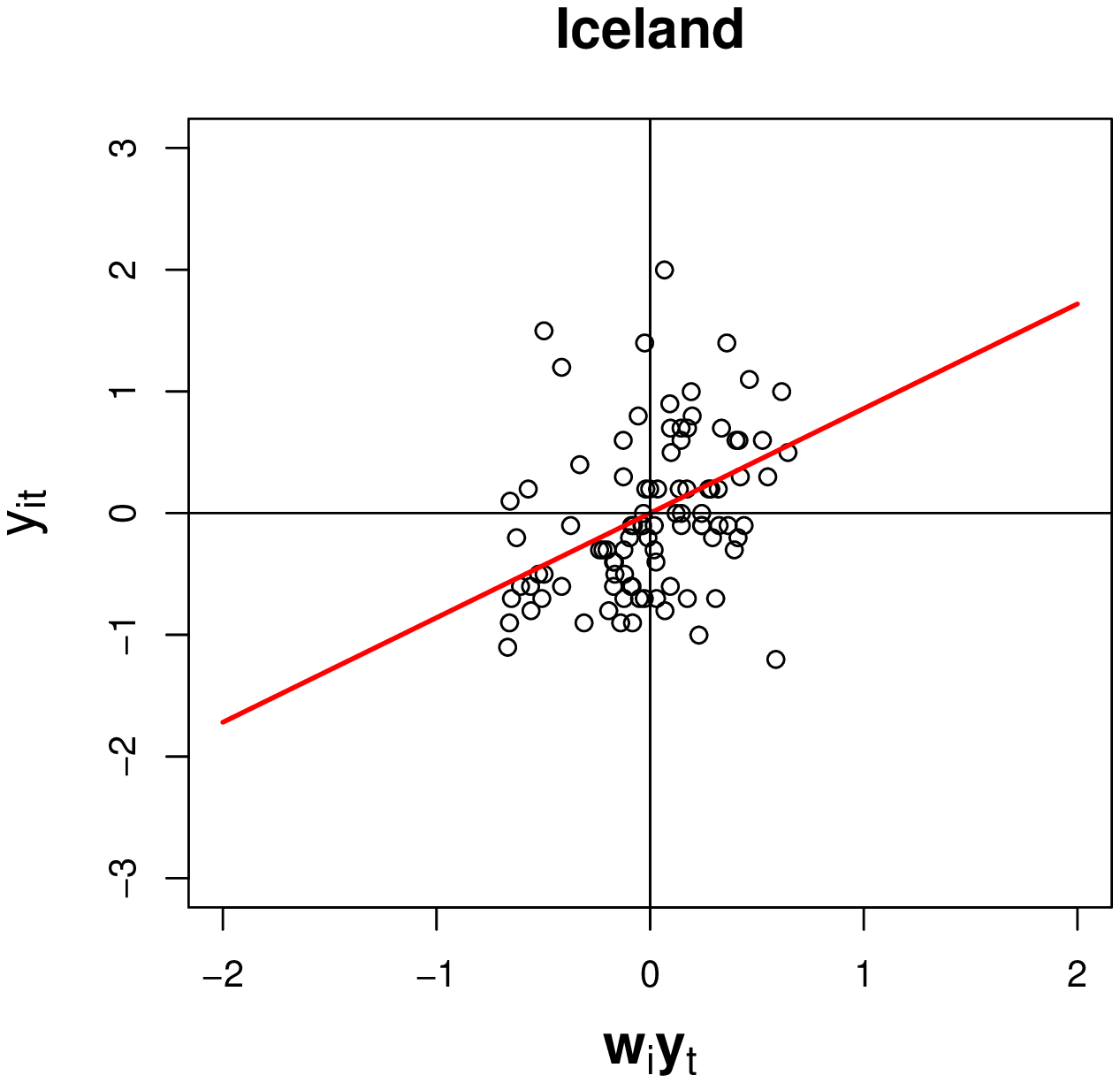}} }
        \mbox{ \resizebox{3.6cm}{!}{\includegraphics{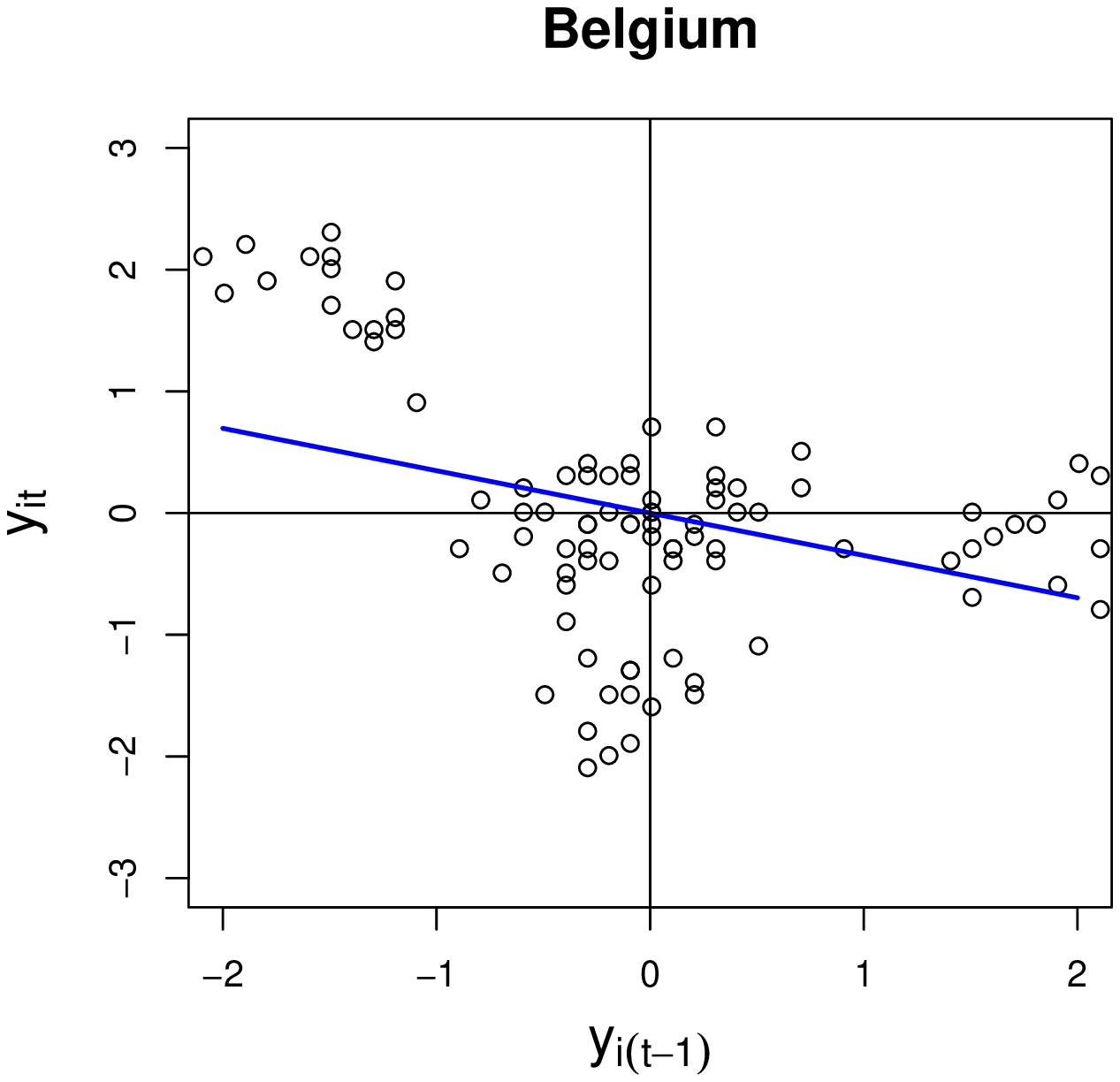}} }
        \mbox{ \resizebox{3.6cm}{!}{\includegraphics{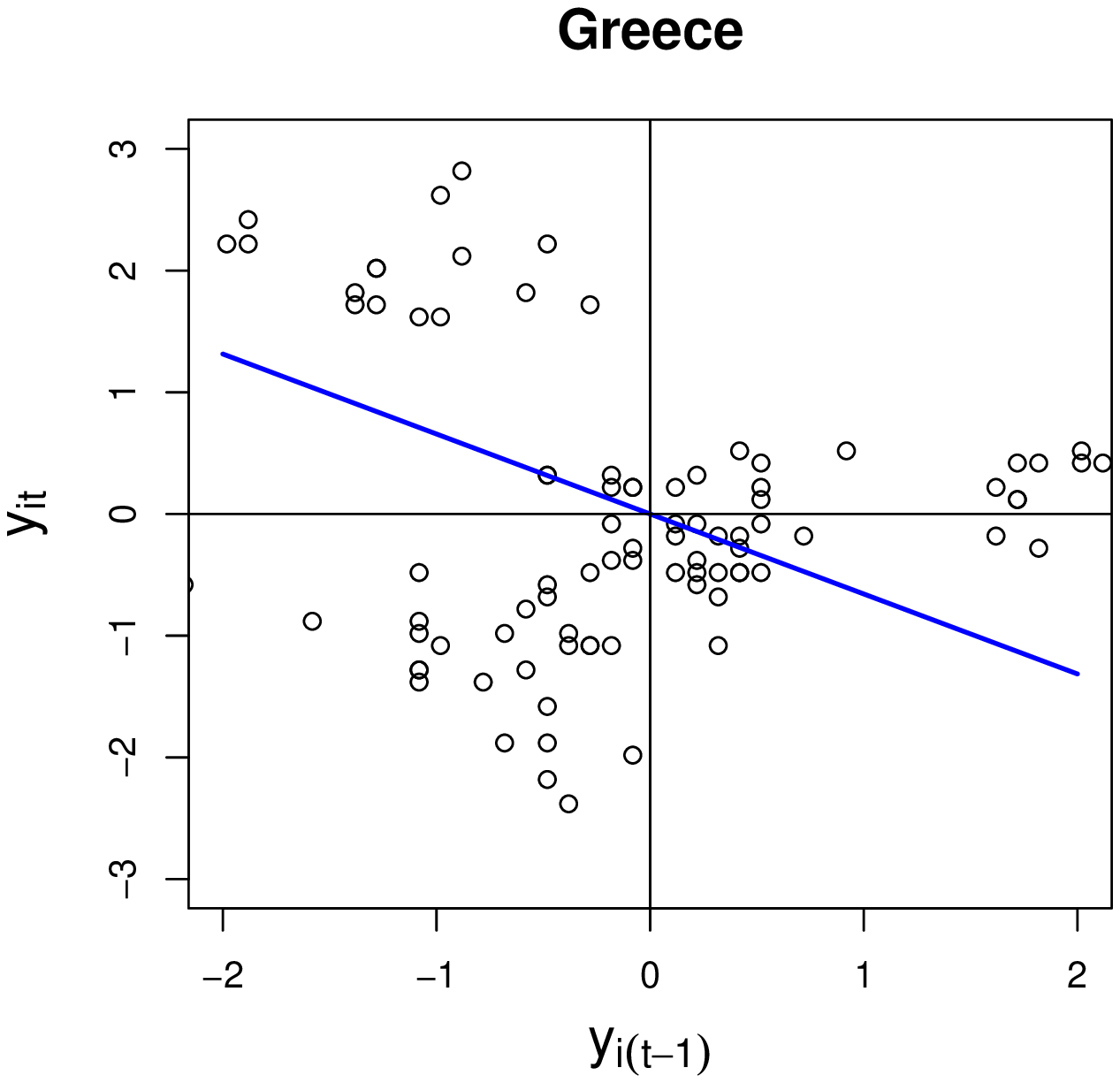}} }
        \mbox{ \resizebox{3.6cm}{!}{\includegraphics{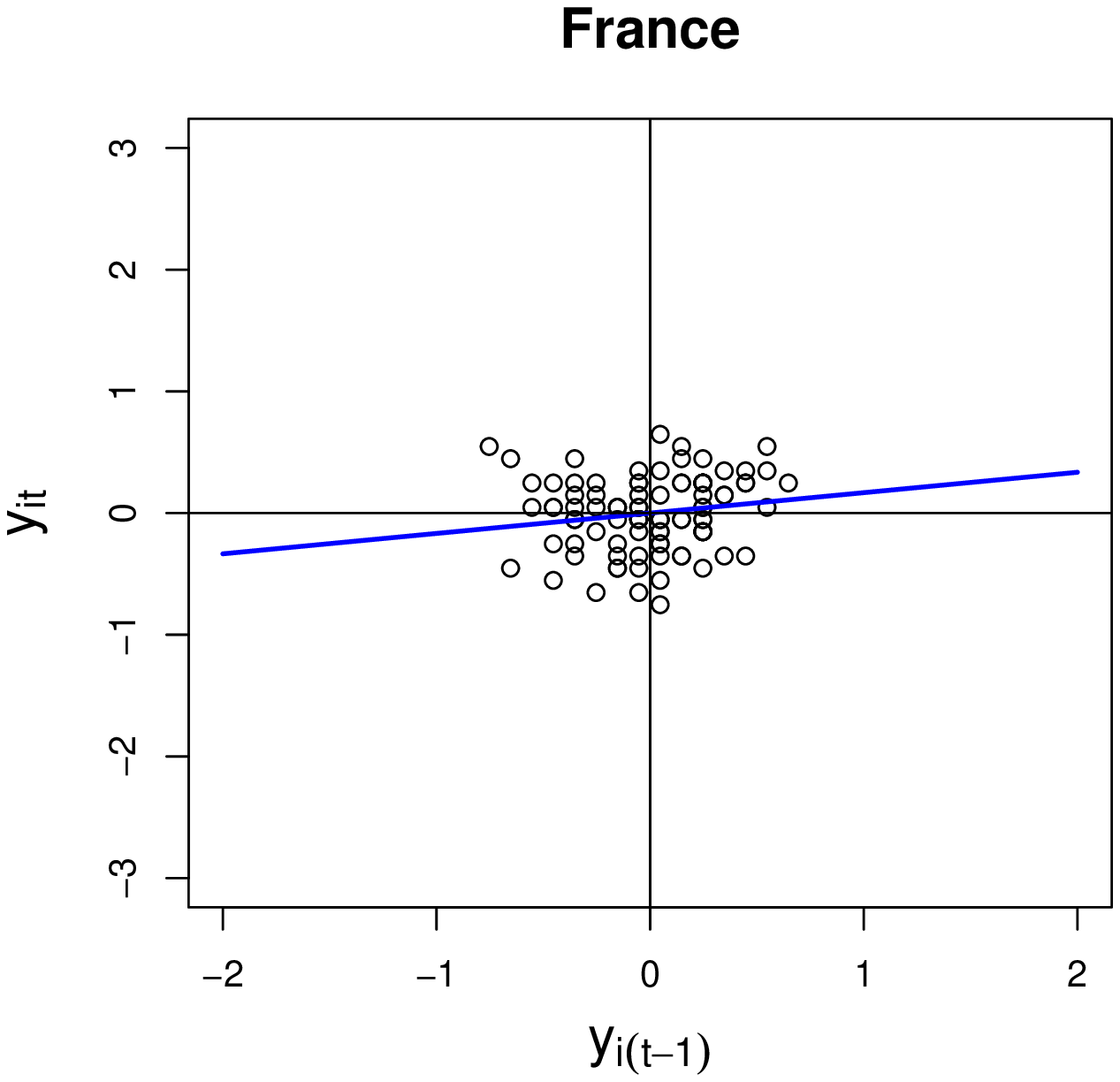}} }
        \mbox{ \resizebox{3.6cm}{!}{\includegraphics{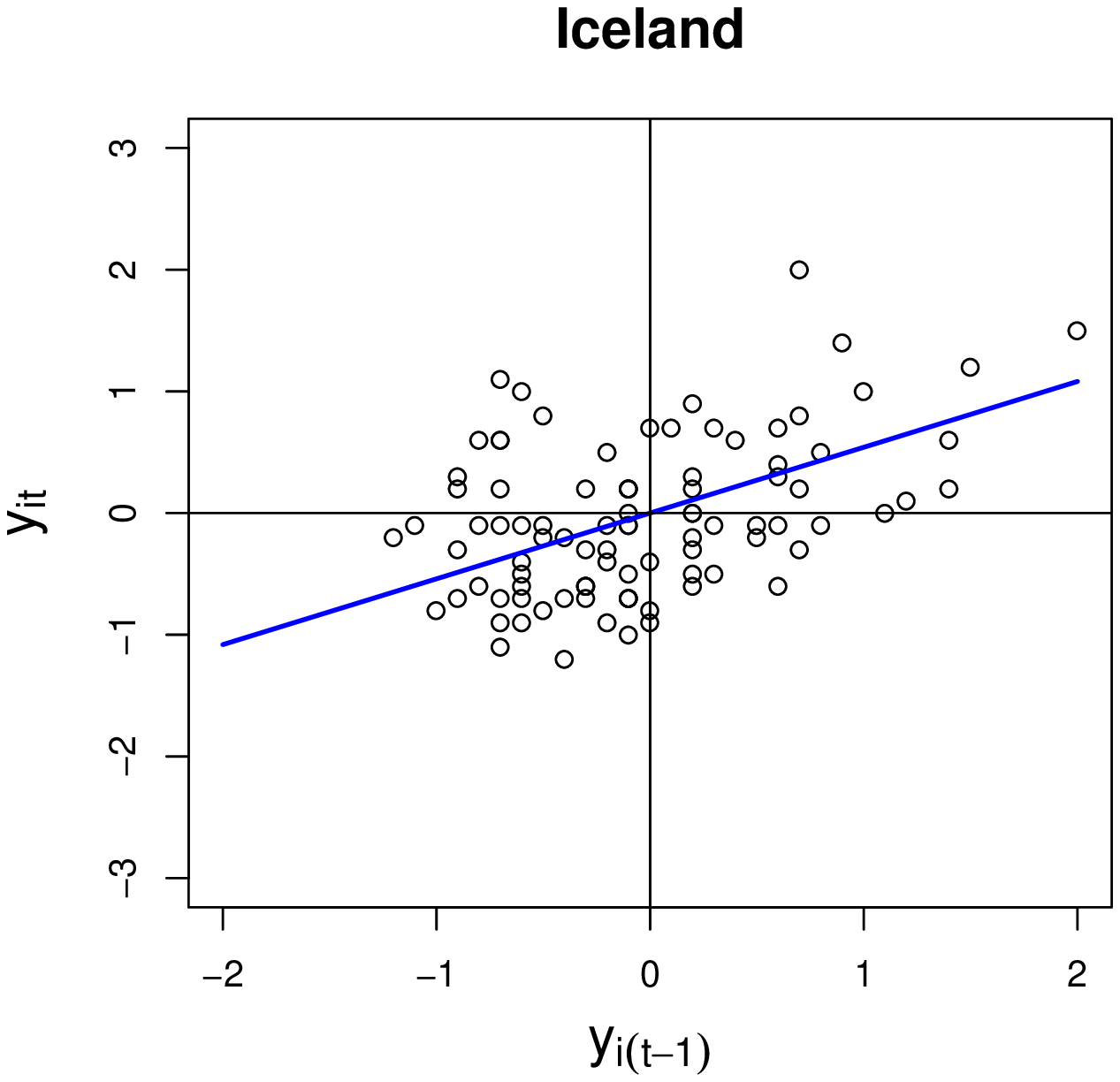}} }
\caption{\label{motivating_example}
Plots of the monthly change rates $y_{i,t}$ of CPI against the spatial
regressor $\bw_i^T \by_t$ (on the top) and the dynamic regressor
$y_{i,t-1}$ (on the bottom) for four EU member states
in 2003-2010.  The superimposed straight lines  were
estimated by the newly proposed method in Section 2.2.}
\end{figure}

The extension to use different scalar coefficients for different locations
is motivated by practical needs.  For example,
we analyze the monthly change rates of the consumer price index (CPI) for the
EU member states over the years 2003-2010. The detailed analysis for this data set
will be presented in section \ref{real data}. Figure \ref{motivating_example} presents the
scatter-plots of the observed data $y_{i,t}$ versus the spatial regressor
$\bw_i^{T}\by_t$ and $y_{i,t-1}$, for some of the EU member states, where $\bw_i^{T}$ is
the $i$-th row vector of the weight matrix  $\bW$ which is taken as
the sample correlation matrix with all the elements on the main diagonal set to be 0.
The superimposed straight lines are the simple regression lines estimated
using the newly proposed method in Section 2.2 below.
It is clear from Figure
\ref{motivating_example} that at least
Greece and Belgium should have a different slope from those of France or Iceland.

\subsection{Generalized Yule-Walker estimation} \label{estimation}
As $\by_t$ occurs on both sides of (\ref{fullmodel}), $\bW\by_t$ and
$\bve_t$ are correlated with each other. Applying least squares method
directly based on regressing $\by_t$ on $(\bW\by_t, \by_{t-1},
\bW\by_{t-1})$ leads to inconsistent estimators. On the
other hand, applying the maximum likelihood estimation requires to
profile a $p\times p$ nuisance parameter matrix $\bSigma_{\bve} =
\var(\bve_t)$, which leads to a complex nonlinear
optimization problem.  Furthermore when $p$ is large in relation to $n$,
the numerical stability is of concern.

We propose below
a new estimation method which applies the least squares method to each individual
row of a Yule-Walker equation. To this end,  let
$
\bSigma_k={\rm{Cov}}(\by_{t+k},\by_t) $ for any $k\ge 0$. Note that we always assume
that $\by_t$ is stationary, see condition A2 and Remark 1 in Section~\ref{asymptotic} below.
Then the Yule-Walker equation below follows from  (\ref{fullmodel})
directly.
\begin{equation}
(\bI-D(\blambda_0)\bW)\bSigma_1=(D({\blambda_1})+ D(\blambda_2)\bW)\bSigma_0, \nonumber
\end{equation}
where $\bI$ is a $p \times p$ identity matrix. The $i$-th row of the above equation is
\begin{equation} \label{equation}
(\be_i^{T}-\lambda_{0i}\bw_i^{T})\bSigma_1=(\lambda_{1i}\be_i^{T}+
\lambda_{2i}\bw_i^{T})\bSigma_0, \quad i=1,\ldots,p, \tag{2}
\end{equation}
where $\bw_i$ is the $i$-th row vector of $\bW$, and $\be_i$ is the unit vector
with the $i$-th element equal to 1.
Note that
(\ref{equation}) is a system of $p$ linear equations with three unknown
parameters $\lambda_{0i}$, $\lambda_{1i}$ and
$\lambda_{2i}$.
Since $E \by_t=\mathbf{0}$, we replace $\bSigma_1$ and
$\bSigma_0$ by the sample (auto)covariance matrices
\[
\widehat\bSigma_1 = {1 \over n} \sum_{t=1}^{n} \by_{t}\by_{t-1}^T
\quad {\rm and} \quad
\widehat\bSigma_0 = {1 \over n} \sum_{t=1}^n \by_t\by_t^T.
\]
We estimate
$(\lambda_{0i},\lambda_{1i},\lambda_{2i})^T$ by
the least squares method, i.e. to solve the minimization problem
\begin{equation}
\min\limits_{\lambda_{0i},\lambda_{1i},\lambda_{2i}}{\|\widehat\bSigma_1^{T}(\be_i-\lambda_{0i}\bw_i)-\widehat\bSigma_0
(\lambda_{1i}\be_i+ \lambda_{2i}\bw_i)\|_2^2}. \nonumber
\end{equation}
The resulting estimators are called generalized Yule-Walker estimators which admits the explicit expression:
\begin{equation} \label{estimator}
(\hat\lambda_{0i},\hat\lambda_{1i},\hat\lambda_{2i})^{T}=(\widehat
\bX_i^{T}\widehat \bX_i)^{-1}\widehat \bX_i^{T}\widehat \bY_i, \tag{3} 
\end{equation}
where
\begin{equation}
\hat \bX_i=(\widehat\bSigma_1^{T}\bw_i, \widehat\bSigma_0\be_i, \widehat\bSigma_0\bw_i)
\quad \text{and} \quad
\hat \bY_i=\widehat\bSigma_1^{T}\be_i.
\nonumber
\end{equation}
More explicitly,
\begin{equation}
\hat \bX_i=\left(\frac{1}{n}\sum_{t=1}^{n}\by_{t-1}(\bw_i^T\by_t),
\frac{1}{n}\sum_{t=1}^{n}\by_{t-1}y_{i,t-1},
\frac{1}{n}\sum_{t=1}^{n}\by_{t-1}(\bw_i^T\by_{t-1})\right),
\quad
\hat \bY_i=\frac{1}{n}\sum_{t=1}^{n}\by_{t-1}y_{i,t}.
\label{hatX} \nonumber
\end{equation}
Then it holds that for $i=1, \cdots, p$,
\begin{equation}
\left(
\begin{array}{c}
\hat\lambda_{0i} \\
\hat\lambda_{1i} \\
\hat\lambda_{2i}
\end{array}
\right)
-
\left(
\begin{array}{c}
\lambda_{0i} \\
\lambda_{1i} \\
\lambda_{2i}
\end{array}
\right)
=(\hat \bX_i^T\hat \bX_i)^{-1}
\left(
\begin{array}{c}
\frac{1}{n}\sum_{t=1}^{n}\by_{t-1}^T(\bw_i^T\by_t) \times \frac{1}{n}\sum_{t=1}^{n}\varepsilon_{i, t}\by_{t-1}  \\
\frac{1}{n}\sum_{t=1}^{n}\by_{t-1}^Ty_{i,t-1} \times \frac{1}{n}\sum_{t=1}^{n}\varepsilon_{i, t}\by_{t-1} \\
\frac{1}{n}\sum_{t=1}^{n}\by_{t-1}^T(\bw_i^T\by_{t-1}) \times \frac{1}{n}\sum_{t=1}^{n}\varepsilon_{i, t}\by_{t-1}
\end{array}
\right). \nonumber
\end{equation}

\subsection{A root-$n$ consistent estimator for large $p$} \label{new_estimation}

When $p/\sqrt{n} \to \infty$, the estimator (\ref{estimator}) admits non-standard
convergence rates (i.e. the rates different from $\sqrt{n}$); see
Theorems 2 and 4 in Section~\ref{asymptotic} below. Note that there are $p$ equations
with only 3 parameters in (\ref{equation}). Hence (\ref{estimator}) can be viewed
as a GMME for an over-determined scenario. The estimation may suffer when the
number of estimation equations increases. See, for example, a similar result in
 Theorem 1 of Chang, Chen and Chen (2015). A further compounding factor is that
the estimation for the covariance matrices $\bSigma_0, \, \bSigma_1$ using
their sample counterparts leads to non-negligible errors even when $n \to \infty$.
Below we propose an alternative estimator which restricts the
number of the estimation equations to be used in order to restore
the $\sqrt{n}$-consistency and the asymptotic normality.

For $i=1,\cdots,p$, put
$
\bX_i=(\bSigma_1^{T}\bw_i, \bSigma_0\be_i, \bSigma_0\bw_i). \nonumber
$
Note that the $k$-th row of $\bX_i$ is
$(\be_k^{T}\bSigma_1^{T}\bw_i, \linebreak \be_k^{T}\bSigma_0\be_i,
\be_k^{T}\bSigma_0\bw_i)$ which is the covariance between $y_{k,t-1}$ and
$(\bw_i^{T}\by_t, \, y_{i,t-1}, \,
\bw_i^{T}\by_{t-1})$. Let
\begin{equation} \label{measure}
\rho_k^{(i)}=\left|\be_k^{T}\bSigma_1^{T}\bw_i\right|+\left|\be_k^{T}\bSigma_0\be_i\right|+
\left|\be_k^{T}\bSigma_0\bw_i\right|, \quad k=1,\cdots,p. \tag{4}
\end{equation}
Then $\rho_k^{(i)}$ may be viewed as a measure for the correlation between
$y_{k,t-1}$ and $(\bw_i^{T}\by_t, y_{i,t-1}, \bw_i^{T}\by_{t-1})^{T}$.
When $\rho_k^{(i)}$ is small, say, close to 0, the $k$-th equation in (\ref{equation})
carries little information on $(\lambda_{0i}, \lambda_{1i}, \lambda_{2i})$.
Therefore as far as  the estimation for $(\lambda_{0i}, \lambda_{1i}, \lambda_{2i})$
is concerned,  we only keep the $k$-th equation in (\ref{equation})
for large $\rho_k^{(i)}$.

Let $\bz_{t-1}^i$ be the $d_i\times 1$ vector consisting of those
$y_{k,t-1}$ corresponding to the $d_i$ largest $\wh\rho_k^{(i)}$ ($1\le k \le p$), where
$\hat\rho_k^{(i)}$ is defined as in (\ref{measure}) but with
$(\bSigma_1, \, \bSigma_0)$ replaced by $(\widehat\bSigma_1 , \, \widehat\bSigma_0)$.
The new estimator is defined as
\begin{equation}\label{new_estimator}
(\wt\lambda_{0i},\, \wt\lambda_{1i},\, \wt\lambda_{2i})^{T}=(\widehat \bZ_i^{T}
\widehat \bZ_i)^{-1}
\widehat \bZ_i^{T}\wt \bY_i, \quad i=1,\cdots,p. \tag{5}
\end{equation}
where
\begin{equation}
\widehat \bZ_i
=\Big(\frac{1}{n}\sum_{t=1}^{n}\bz^i_{t-1}(\bw_i^T\by_t),\;
\frac{1}{n}\sum_{t=1}^{n}\bz^i_{t-1}y_{i,t-1},\;
\frac{1}{n}\sum_{t=1}^{n}\bz^i_{t-1}(\bw_i^T\by_{t-1})\Big), \tag{6}
\label{zi}
\end{equation}
and
\begin{equation}
\wt \bY_i=\frac{1}{n}\sum_{t=1}^{n}\bz^i_{t-1}y_{i,t}. \nonumber
\end{equation}
Now it holds that
\begin{equation}
\left(
\begin{array}{c}
\wt\lambda_{0i} \\
\wt\lambda_{1i} \\
\wt\lambda_{2i}
\end{array}
\right)
-
\left(
\begin{array}{c}
\lambda_{0i} \\
\lambda_{1i} \\
\lambda_{2i}
\end{array}
\right)
=(\widehat \bZ_i^{T}\widehat \bZ_i)^{-1}
\widehat \bZ_i^{T}
\left(
\begin{array}{c}
\frac{1}{n}\sum_{t=1}^{n}\varepsilon_{i, t}\bz^i_{t-1}  \\
\frac{1}{n}\sum_{t=1}^{n}\varepsilon_{i, t}\bz^i_{t-1} \\
\frac{1}{n}\sum_{t=1}^{n}\varepsilon_{i, t}\bz^i_{t-1}
\end{array}
\right).
\nonumber
\end{equation}
Theorem \ref{theorem3} in Section 3 below shows the asymptotic normality
of the above estimator provided that the number of estimation equations used
satisfies condition
$d_i = o(\sqrt{n})$.

\section{Theoretical properties}\label{asymptotic}

We introduce some notations first. For a $p \times 1$ vector $\bv =(v_1, \cdots, v_p)^T$,
$\|\bv\|_2=\sqrt{\sum_{i=1}^p v_i^2}$ is the Euclidean norm,
$\|\bv\|_1=\sum_{i=1}^p \left|v_i\right|$ is the $L_1$ norm. For a
matrix $\bH =(h_{ij})$, $\|\bH\|_F = \sqrt{tr(\bH^{T}\bH)}$ is the Frobenius norm,
$\|\bH\|_2=\sqrt{\lambda_{\max}(\bH^{T}\bH)}$ is the operator norm, where
$\lambda_{\max}(\cdot)$ is the largest eigenvalue of a matrix. We denote
by $\left|\bH\right|$ the matrix $(\left|
h_{ij}\right|)$ which is a matrix of the same size as $\bH$ but with the
$(i,j)$-th element $h_{ij}$ replaced
by $|h_{ij}|$.
 Note the determinant of $\bH$ is denoted by $\det(\bH)$.
A strictly stationary process $\{ \by_t\}$ is  $\alpha$-mixing if
\begin{equation} \label{mixing}
\alpha(k)\equiv \sup_{A \in
\mathcal{F}_{-\infty}^{0}, B \in
\mathcal{F}_{k}^{\infty}}\big|P(A)P(B)-P(AB)\big|
\to 0, \quad {\rm as} \;\; k \to \infty, \tag{7}
\end{equation}
where $\mathcal{F}_{i}^{j}$ denotes the $\sigma$-algebra generated by
$\{\by_t, i \le t \le j\}$. See,  e.g., Section 2.6 of
Fan and Yao (2003) for a compact review of $\alpha$-mixing processes.

Let
$\bS(\blambda_0) \equiv \bI-D(\blambda_0)\bW$ be invertible. It follows from (\ref{fullmodel})
that
\begin{equation}\label{reducemodel}
\by_t=\bA\by_{t-1} + \bS^{-1}(\blambda_0)\bve_t, \nonumber
\end{equation}
where $\bA=\bS^{-1}(\blambda_0)(D({\blambda_1})+ D(\blambda_2)\bW)$.
Some regularity conditions are now in order.
\begin{itemize}
\item[A1.] The spatial weight matrix $\bW$ is known with zero main diagonal
elements; $\bS(\blambda_0)$ is invertible.
\item[A2.]
(a) The disturbance $\bve_t$ satisfies
\[
\mathrm{Cov}(\by_{t-1},\bve_t)=0.
\] \\
(b) The process $\{ \by_t \}$ in model (\ref{fullmodel}) is strictly
stationary and $\alpha$-mixing with $\alpha(k)$, defined in (\ref{mixing}),
satisfying
\[
\sum_{k=1}^{\infty}\alpha(k)^{\frac{\gamma}{4+\gamma}}< \infty,
\]
for some constant $\gamma>0$.
\\
(c) For 
$\gamma >0$ specified in (b) above,
\[
\sup_{p} \mathrm{E}\left|\bw_i^T\bSigma_0\by_t\right|^{4+\gamma} < \infty, \quad \sup_{p} \mathrm{E}\left|\bw_i^T\bSigma_1\by_t\right|^{4+\gamma} < \infty, \quad \sup_{p} \mathrm{E}\left|\be_i^T\bSigma_0\by_t\right|^{4+\gamma} < \infty,
\]
\[
\sup_{p} \mathrm{E}\left|\bw_i^T\by_t\right|^{4+\gamma} < \infty, \quad \sup_{p} \mathrm{E}\left|\be_i^T\by_t\right|^{4+\gamma} < \infty,
\]
where $\bw_i$ denotes the $i$-th row of $\bW$. The diagonal elements of
$\bV_i$ defined in (\ref{V}) are bounded uniformly in $p$.
\item[A3.]
The rank of matrix
$ 
(\bSigma_1^{T}\bw_i, \bSigma_0\be_i, \bSigma_0\bw_i)$
is equal to 3. 
\end{itemize}

\noindent
{\bf Remark 1}. Condition A1 is standard for spatial econometric models. Condition A3 ensures
that $\la_{0i}, \la_{1i}$ and $\la_{2i}$ are identifiable in
(\ref{equation}). Condition A2(c) limits the dependence
across different spatial locations.
It is implied by, for example, the conditions imposed in Yu et al. (2008).
Lemma \ref{lemma1} in the Appendix shows that
 Condition A2 holds with $\gamma=4$ under conditions A1 and B1 -- B3 below.
Note that conditions B1--B3 are often directly imposed in the spatial
econometrics literature including, for example,
Lee and Yu (2010a), and Yu et al. (2008).
\begin{itemize}
\item[B1.] The errors $\varepsilon_{i,t}$
 are $i.i.d$ across $i$ and $t$ with
$\mathrm{E}(\varepsilon_{i,t})=0$, $\var(\varepsilon_{i,t})=\sigma_0^2$, and
$\mathrm{E}\left|\varepsilon_{i,t}\right|^{4+\gamma} < \infty$.
The density function of $\varepsilon_{i,t}$ exists.
\item[B2.] The row and column sums of $\left|\bW\right|$ and
$\left|\bS^{-1}(\blambda_0)\right|$ are bounded uniformly in $p$.
\item[B3.] The row and column sums of
$\sum_{j=0}^{\infty}\left|\bA^{j}\right|$ are bounded uniformly in $p$.
\end{itemize}

Now we are ready to present  the asymptotic properties for
$(\hat\lambda_{0i},\hat\lambda_{1i},\hat\lambda_{2i})^{T}$, $i=1,\ldots,p$,
with fixed $p$ and $n \to \infty$ first, and then $p \to \infty$ and $n \to \infty$.

\subsection{Asymptotics for fixed $p$}
For $i=1,\ldots,p$, let
\[
\bSigma_{\by, \varepsilon_{i}}(j)=\cov(\by_{t-1+j}\varepsilon_{i,t+j},
\by_{t-1}\varepsilon_{i,t}), \quad \quad j=0,1,2,\cdots,
\]
\[
\bSigma_{\by, \varepsilon_{i}}=\bSigma_{\by,
\varepsilon_{i}}(0)+\sum_{j=1}^{\infty}\left[\bSigma_{\by,
\varepsilon_{i}}(j)+\bSigma_{\by, \varepsilon_{i}}^T(j)\right],
\]
\begin{equation} \label{V}
\bV_i=
\left(
\begin{array}{ccc}
\bw_i^T\bSigma_1\bSigma_1^T\bw_i & \bw_i^T\bSigma_1\bSigma_0 \be_i & \bw_i^T\bSigma_1\bSigma_0\bw_i \\
\bw_i^T\bSigma_1\bSigma_0 \be_i  & \be_i^T\bSigma_0\bSigma_0 \be_i & \be_i^T\bSigma_0\bSigma_0\bw_i \\
\bw_i^T\bSigma_1\bSigma_0\bw_i   & \be_i^T\bSigma_0\bSigma_0\bw_i  & \bw_i^T\bSigma_0\bSigma_0\bw_i
\end{array}
\right), \tag{8}
\end{equation}
and
\begin{equation} \label{U}
\bU_i=
\left(
\begin{array}{ccc}
\bw_i^T\bSigma_1\bSigma_{\by, \varepsilon_{i}}\bSigma_1^T\bw_i & \bw_i^T\bSigma_1\bSigma_{\by, \varepsilon_{i}}\bSigma_0 \be_i & \bw_i^T\bSigma_1\bSigma_{\by, \varepsilon_{i}}\bSigma_0\bw_i \\
\bw_i^T\bSigma_1\bSigma_{\by, \varepsilon_{i}}\bSigma_0 \be_i  & \be_i^T\bSigma_0\bSigma_{\by, \varepsilon_{i}}\bSigma_0 \be_i & \be_i^T\bSigma_0\bSigma_{\by, \varepsilon_{i}}\bSigma_0\bw_i \\
\bw_i^T\bSigma_1\bSigma_{\by, \varepsilon_{i}}\bSigma_0\bw_i   & \be_i^T\bSigma_0\bSigma_{\by, \varepsilon_{i}}\bSigma_0\bw_i  & \bw_i^T\bSigma_0\bSigma_{\by, \varepsilon_{i}}\bSigma_0\bw_i
\end{array}
\right). \tag{9}
\end{equation}
\begin{theorem}\label{theorem1}
Let conditions A1 -- A3 hold and $p\ge 1$ be fixed. Then as $n \to \infty$,
it holds that
\begin{equation}
\sqrt{n}
\left(
\left(
\begin{array}{c}
\hat\lambda_{0i} \\
\hat\lambda_{1i} \\
\hat\lambda_{2i}
\end{array}
\right)
-
\left(
\begin{array}{c}
\lambda_{0i} \\
\lambda_{1i} \\
\lambda_{2i}
\end{array}
\right)
\right) \xrightarrow{d} N(0,\bV_i^{-1}\bU_i\bV_i^{-1}), \qquad i=1,\ldots,p, \nonumber
\end{equation}
where $\bV_i$ and $\bU_i$ are given in $(\ref{V})$ and $(\ref{U})$.
\end{theorem}

\subsection{Asymptotics with diverging $p$} 
\label{asylargep}

When $p$ diverges together with $n$, $\bU_i, \bV_i$ in (\ref{U}) and
(\ref{V}) are no longer constant matrices.
Let $\bU_i^{-\frac{1}{2}}$ be a matrix such that $(\bU_i^{-\frac{1}{2}})^2 = \bU_i^{-1}$.

\begin{theorem}\label{theorem2}
Let condition A1 -- A3 hold.
\begin{itemize}
\item[(\romannumeral1)]
 As $n \to \infty$, $p \to \infty$ and $p=o(\sqrt{n})$,
\begin{eqnarray} \label{result2}
\sqrt{n} \bU_i^{-\frac{1}{2}} \bV_i
\left(
\left(
\begin{array}{c}
\hat\lambda_{0i} \\
\hat\lambda_{1i} \\
\hat\lambda_{2i}
\end{array}
\right)
-
\left(
\begin{array}{c}
\lambda_{0i} \\
\lambda_{1i} \\
\lambda_{2i}
\end{array}
\right)
\right) \xrightarrow{d} N(0,\bI_3), \qquad i=1,\ldots,p. \nonumber
\end{eqnarray}
\item[(\romannumeral2)] 
 As $n \to \infty$, $p \to \infty$,  $\sqrt{n}=O(p)$ and $p=o(n)$,
\begin{equation}
\left\|
\left(
\begin{array}{c}
\hat\lambda_{0i} \\
\hat\lambda_{1i} \\
\hat\lambda_{2i}
\end{array}
\right)
-
\left(
\begin{array}{c}
\lambda_{0i} \\
\lambda_{1i} \\
\lambda_{2i}
\end{array}
\right)\right\|_2=O_p\left(\frac{p}{n}\right), \qquad i=1,\ldots,p. \nonumber
\end{equation}
\end{itemize}
\end{theorem}

Theorem 2 indicates that the standard root-$n$ convergence rate prevails as long as
$p= o(\sqrt{n})$. However the convergence rate may be slower when $p$ is of  higher orders
than $\sqrt{n}$.
Theorem 2 presents the convergence rates for the $L_2$ norm
of the estimation errors. The rates also hold for the $L_1$ norm of the errors as well.
Corollary 1 consider the estimation errors over $p$ locations together, for which
we have established the result for $L_1$ norm only.
\begin{corollary} \label{corollary1}
Let condition A1 hold, and condition A2 and A3  hold for all $i=1, \cdots, p$.
Then as $n \to \infty$ and $p \to \infty$, it holds that
\begin{equation}
{1 \over p} \sum_{i=1}^p \left\|
\left(
\begin{array}{c}
\hat\lambda_{0i} \\
\hat\lambda_{1i} \\
\hat\lambda_{2i}
\end{array}
\right)
-
\left(
\begin{array}{c}
\lambda_{0i} \\
\lambda_{1i} \\
\lambda_{2i}
\end{array}
\right)\right\|_1 =
\begin{cases}
O_p(\frac{1}{\sqrt{n}}) \quad & {\rm if} \;\; \frac{p}{\sqrt{n}}=O(1), \\
O_p(\frac{p}{n}) & {\rm if} \;\; \frac{p}{\sqrt{n}} \to \infty \;\; {\rm and} \;\; \frac{p}{n}=o(1).
\end{cases} \nonumber
\end{equation}
\end{corollary}


\askip

To derive the asymptotic properties of the estimators defined in (\ref{new_estimator}),
we introduce some new notation.
For $i=1,\ldots,p$, let
\[
\bSigma_0^i =\cov(\by_t, \bz_{t}^i), \quad \bSigma_1^i =\cov(\by_t, \bz_{t-1}^i),
\]
\[
\bSigma_{\bz^{i},
\varepsilon_{i}}(j)=\cov(\bz_{t-1+j}^i\varepsilon_{i,t+j},
\bz_{t-1}^i\varepsilon_{i,t}), \quad \quad j=0,1,2,\cdots,
\]
and
\[
\bSigma_{\bz^i, \varepsilon_{i}}=\bSigma_{\bz^i, \varepsilon_{i}}(0)+\sum_{j=1}^{\infty}\left[\bSigma_{\bz^i, \varepsilon_{i}}(j)+\bSigma_{\bz^i, \varepsilon_{i}}^T(j)\right].
\]
Let
\begin{equation} \label{V*}
\bV_i^*=
\left(
\begin{array}{ccc}
\bw_i^T\bSigma_1^i (\bSigma_1^i )^T\bw_i &
\bw_i^T\bSigma_1^i(\bSigma_0^i )^{T} \be_i &
\bw_i^T\bSigma_1^i (\bSigma_0^i )^{T}\bw_i \\
\bw_i^T\bSigma_1^i (\bSigma_0^i )^{T} \be_i  &
\be_i^T\bSigma_0^i(\bSigma_0^i )^{T} \be_i &
\be_i^T\bSigma_0^i(\bSigma_0^i)^{T}\bw_i \\
\bw_i^T\bSigma_1^i (\bSigma_0^i )^{T}\bw_i   &
\be_i^T\bSigma_0^i (\bSigma_0^i )^{T}\bw_i  &
\bw_i^T\bSigma_0^i (\bSigma_0^i )^{T}\bw_i
\end{array}
\right), \tag{10}
\end{equation}
and
\begin{equation} \label{U*}
 \bU_i^*=
\left(
\begin{array}{ccc}
\bw_i^T\bSigma_1^i \bSigma_{\bz^i, \varepsilon_{i}}(\bSigma_1^i )^T\bw_i &
\bw_i^T\bSigma_1^i \bSigma_{\bz^i, \varepsilon_{i}}(\bSigma_0^i )^{T} \be_i &
\bw_i^T\bSigma_1^i \bSigma_{\bz^i, \varepsilon_{i}}(\bSigma_0^i )^{T}\bw_i \\
\bw_i^T\bSigma_1^i \bSigma_{\bz^i, \varepsilon_{i}}(\bSigma_0^i )^T \be_i  &
\be_i^T\bSigma_0^i \bSigma_{\bz^i, \varepsilon_{i}}(\bSigma_0^i )^T \be_i &
\be_i^T\bSigma_0^i \bSigma_{\bz^i, \varepsilon_{i}}(\bSigma_0^i )^T\bw_i \\
\bw_i^T\bSigma_1^i \bSigma_{\bz^i, \varepsilon_{i}}(\bSigma_0^i )^T\bw_i   &
\be_i^T\bSigma_0^i \bSigma_{\bz^i, \varepsilon_{i}}(\bSigma_0^i )^T\bw_i  &
\bw_i^T\bSigma_0^i \bSigma_{\bz^i, \varepsilon_{i}}(\bSigma_0^i )^T\bw_i
\end{array}
\right). \tag{11}
\end{equation}

Theorem \ref{theorem3} below indicates that the estimators defined
in (\ref{new_estimator}) are asymptotically normal with the standard
$\sqrt{n}$-rate as long as $d_i = o(\sqrt{n})$.
Note that it does not impose any conditions directly on the size of $p$.
\begin{itemize}
\item[A4.] (a) For $\gamma>0$ specified in A2(b),
 \[
\sup_{p} \mathrm{E}\left|\bw_i^T\bSigma_0^i \bz_t^i\right|^{4+\gamma} < \infty,
\quad \sup_{p} \mathrm{E}\left|\bw_i^T\bSigma_1^i \bz_t^i\right|^{4+\gamma} < \infty,
\quad \sup_{p} \mathrm{E}\left|\be_i^T\bSigma_0^i \bz_t^i\right|^{4+\gamma} < \infty,
\]
\[
\sup_{p} \mathrm{E}\left|\bw_i^T\by_t\right|^{4+\gamma} < \infty, \quad \sup_{p} \mathrm{E}\left|\be_i^T\by_t\right|^{4+\gamma} < \infty.
\]
and the diagonal elements of $\bV_i^*$ defined in (\ref{V*}) are bounded uniformly in $p$. \\
(b) The rank of matrix $\mathrm{E}\{ \widehat \bZ_i \}$ is equal to 3, where $\widehat \bZ_i$
is defined in (\ref{zi}).
\end{itemize}

\begin{theorem}\label{theorem3}
Let conditions A1, A2(a,b) and A4 hold. As $n \to \infty$, $ p \to \infty$ and $d_i=o(\sqrt{n})$, it holds that
\begin{eqnarray} \label{result3}
\sqrt{n} (\bU_i^*)^{-\frac{1}{2}} \bV_i^*
\left(
\left(
\begin{array}{c}
\wt\lambda_{0i} \\
\wt\lambda_{1i} \\
\wt\lambda_{2i}
\end{array}
\right)
-
\left(
\begin{array}{c}
\lambda_{0i} \\
\lambda_{1i} \\
\lambda_{2i}
\end{array}
\right)
\right) \xrightarrow{d} N(0,\bI_3),\qquad i=1,\ldots,p, \nonumber
\end{eqnarray}
where $\bV_i^*$ and $\bU_i^*$ are given in $(\ref{V*})$ and $(\ref{U*})$.
\end{theorem}

The key assumption of Theorem 2 is A2(c), which decides the fact that the effect of the dimensionality $p$ only comes from $E_1$ in equation (\ref{error_decom}) in the Appendix. We can relax this assumption by allowing $E_2$ to be affected by $p$ as well. Under the new relaxed assumption, we may obtain a better convergent rate of estimator (\ref{estimator}) by making use of the fact that (\ref{estimator}) is invariant if we divide both the numerator and denominator by the same number, for example, a number relating to $p$. This will be presented in Theorem 4. We propose the new relaxed assumption:
\begin{itemize}
\item[A5.]
For $\gamma >0$ specified in A2(b),
\[
\max\Big\{\sup_{p} \mathrm{E}\left|\bw_i^T\bSigma_0\by_t\right|^{4+\gamma}, \quad \sup_{p} \mathrm{E}\left|\bw_i^T\bSigma_1\by_t\right|^{4+\gamma}, \quad \sup_{p} \mathrm{E}\left|\be_i^T\bSigma_0\by_t\right|^{4+\gamma}\Big\}=O(s_0(p)).
\]
\[
\max\Big\{\sup_{p} \mathrm{E}\left|\bw_i^T\by_t\right|^{4+\gamma}, \quad \sup_{p} \mathrm{E}\left|\be_i^T\by_t\right|^{4+\gamma} \Big\}=O(s_1(p)).
\]
and the diagonal elements of $\bV_i$ defined in (\ref{V}) is in the order of $s_2(p)$, where $s_0(p)$, $s_1(p)$ and $s_2(p)$ are numbers relating to $p$.
\end{itemize}

Denote $C$ as a constant. When the number of nonzero elements (or elements bounded away from zero) in $\bw_i$ increases with $p$ but is $o(p)$, we may have $s_1(p)=o(\min\{s_0(p), s_2(p)\})$. Simulation scenario 2 is under this case. When there are only finite number of nonzero elements (or elements bounded away from zero) in $\bw_i$, we might have $s_1(p) \asymp C$, which is the case of simulation scenario 1. The reason we assume the diagonal elements of $\bV_i$ defined in (\ref{V}) are in the order of $s_2(p)$ is because we can treat $\bw_i^T\bSigma_1\bSigma_1^T\bw_i, \be_i^T\bSigma_0\bSigma_0 \be_i, \bw_i^T\bSigma_0\bSigma_0\bw_i$ as the second moments of three random variables $\bw_i^T\bSigma_1\bx, \be_i^T\bSigma_0\bx$ and $\bw_i^T\bSigma_0\bx$ respectively, where the $p \times 1$ random vector $\bx$ has mean 0 and covariance matrix $\bI_p$.

\begin{theorem}\label{theorem4}
Let conditions A1, A2(a,b), A3 and A5 hold. As $n \to \infty$, $ p \to \infty$, if $\frac{ps_1(p)}{s_2(p)}=o(n)$ and $s_0^{1/2}(p)=O(ps_1^{1/2}(p)s_2(p))$, it holds that
\begin{eqnarray} \label{result4}
\left\|
\left(
\begin{array}{c}
\wh\lambda_{0i} \\
\wh\lambda_{1i} \\
\wh\lambda_{2i}
\end{array}
\right)
-
\left(
\begin{array}{c}
\lambda_{0i} \\
\lambda_{1i} \\
\lambda_{2i}
\end{array}
\right)
\right\|_2= O_p\Big(\max \Big\{ \frac{p s_1^{3/4}(p)}{n s_2(p)},\frac{s_0^{1/4}(p)}{\sqrt{n}s_2(p)} \Big\}\Big). \nonumber
\end{eqnarray}
\end{theorem}

Let us consider some examples. (1) When $s_0(p) \asymp p$, $s_1(p) \asymp C$ and $s_2(p) \asymp p$, the convergence rate is $\max \Big\{ \frac{1}{n},\frac{1}{\sqrt{n}p^{3/4}} \Big\}$. (2) When $s_0(p) \asymp p$, $s_1(p) \asymp \sqrt{p}$ and $s_2(p) \asymp p$, if $p=o(n^2)$, the convergence rate is $\max \Big\{ \frac{p^{3/8}}{n},\frac{1}{\sqrt{n}p^{3/4}} \Big\}$. (3) When $s_0(p) \asymp C$, $s_1(p) \asymp C$ and $s_2(p) \asymp C$, if $p=o(n)$, the convergence rate is $\max \Big\{ \frac{p}{n},\frac{1}{\sqrt{n}} \Big\}$, which corresponds with Theorem 2. Theorem 4 indicates that under different situations of $s_0(p)$, $s_1(p)$ and $s_2(p)$, we may obtain different convergence rates. These observations are illustrated by simulation examples in section 4.

\section{Simulation study} \label{simulation}

To examine the finite sample performance of the proposed estimation methods, we conduct some simulation under different scenarios.

\subsection{Scenario 1}
$\lambda_{0i}$, $\lambda_{1i}$ and $\lambda_{2i}$ are generated from $U(-0.6,0.6)$. The spatial weight matrix $\bW$ used is a block diagonal matrix formed by a $\sqrt{p} \times \sqrt{p}$ row-normalized matrix $\bW^*$. We construct $\bW^*$ such that the first four sub-diagonal elements are all 1 and the rest elements are all 0 before normalizing. This kind of $\bW$ corresponds to the pooling of $\sqrt{p}$ separate districts with similar neighboring structures in each district, see Lee and Yu (2013). The error $\varepsilon_{i,t}$ are independently generated from $N(0,\sigma_i^2)$, where we generate each $\sigma_i$ from $U(0.5,1.5)$.

For all scenarios, we generate data from (2.1) with different settings for $n$ and $p$. We apply the proposed estimation method (2.3) and (2.5) (with $d_i=\min{(p, n^{10/21})}$) and report the mean absolute errors:
\begin{equation}\label{mse}
\MAE(i) = {1\over 3} \sum_{j=0}^2 |\wh \la_{ji} - \la_{ji}|, \qquad
\MAE = {1 \over p} \sum_{i=1}^p \MAE(i). \nonumber
\end{equation}
We replicate each setting 500 times.

\begin{figure}[ht]
	\centering
	\mbox{ \resizebox{11cm}{!}{\includegraphics{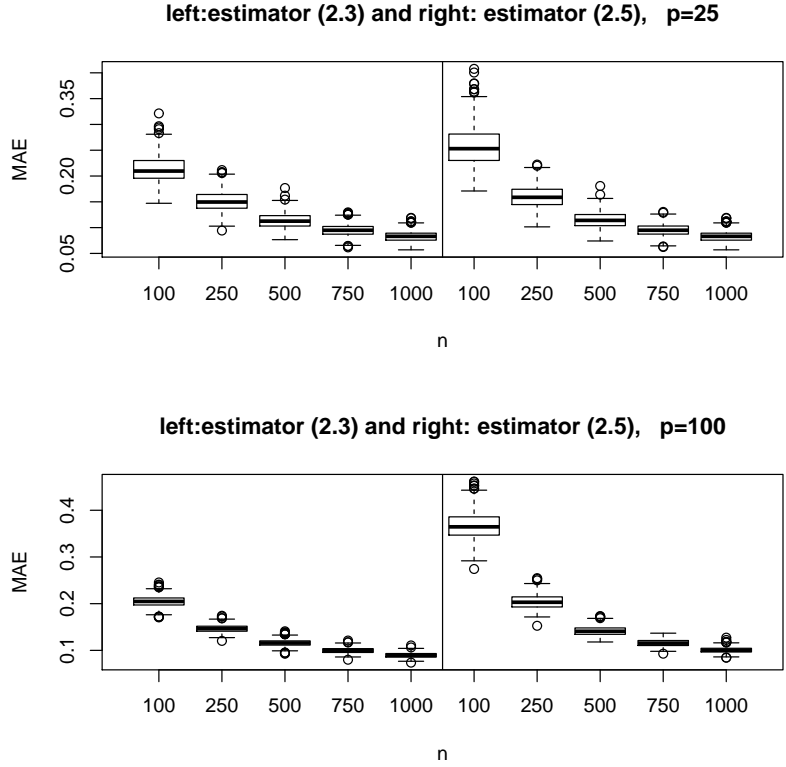}} }
\caption{\label{figure1} Boxplots of MAE for estimator (2.3) (left panels) and
estimator (2.5) (right panels) with $p=25 $ (top panels) and 100 (bottom panels), $n=100,\,250,\,500,\,750,\,1000$ for scenario 1.}
\end{figure}

Figure \ref{figure1} depicts two boxplots of MAE with $p$ equals to, respectively, 25 and 100. As the sample size $n$ increases from 100, 250, 500, 750 to 1000, MAE decreases for both methods.

\begin{figure}[ht]
	\centering
	\mbox{ \resizebox{11cm}{!}{\includegraphics{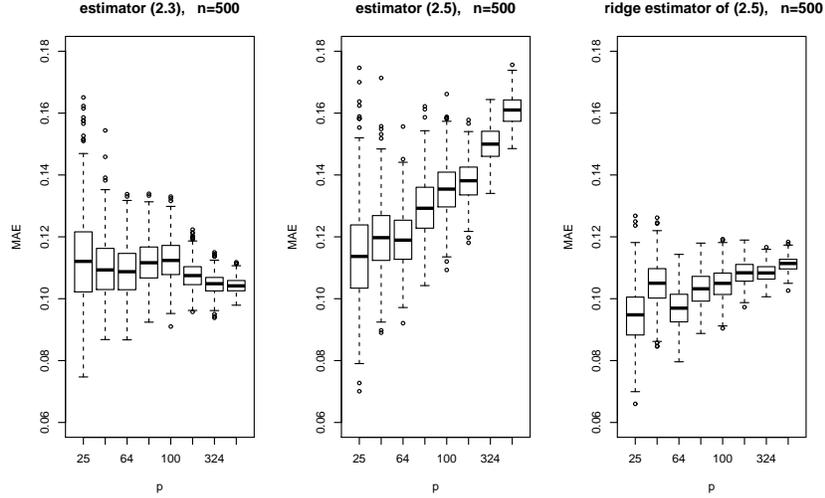}} }
\caption{\label{figure2} Boxplots of MAE of the original estimator (2.3) (the left panel),
the root $n$ consistent estimator (2.5) (the middle panel), and the estimator (2.5) after adding ridge penalty (the right panel)  with $n=500$ and $p=25,49,64,81,100,169,324,529$ for scenario 1.}
\end{figure}

Figure \ref{figure2} depicts the boxplots of the MAE for the
original estimator (2.3), the
root $n$ consistent estimator (2.5), and the estimator (2.5)
with the ridge penalty, where we choose the ridge tuning
parameter to be $C \times \frac{p}{n}$ in order to avoid the nearly
singularity problem of $\hat\bZ_i^T\hat\bZ_i$,  and $C$ is chosen via cross
validation. With $n=500$, the dimension $p$ is set at 25,49,64,81,100,169,324 and 529
respectively. The MAE for (2.3) remains about the same level as $p$ increases; see
the panel on the left in Figure \ref{figure2}.  This
is in line with the asymptotic result of  Theorem 4 when, for example, $s_1(p) \asymp
C$, $s_0(p) \asymp p $ and $ s_2(p) \asymp p$. In contrast, the MAE for estimator (2.5)
increases sharply when $p$ increases; see the panel in the middle. This
is due to the fact that $\hat\bZ_i^T\hat\bZ_i$ is nearly singular for large $p$.
Adding a ridge in the estimator certainly mitigates the deterioration
when $p$ increases; see the panel on the right in Figure \ref{figure2}.

\subsection{Scenario 2}
$\lambda_{0i}$, $\lambda_{1i}$ and $\lambda_{2i}$ are generated from
$U(-0.6,0.6)$. The spatial weight matrix $\bW$ is constructed as follows.
First, we construct a $\sqrt{p} \times \sqrt{p}$ row-normalized matrix
$\bW^*$, where $\bW^*$ is chosen such that the first two sub-diagonal
elements are all 1 and the rest elements are all 0 before normalizing.
Then we treat $\bW$ as a $\sqrt{p} \times \sqrt{p}$ block matrix and put
$\bW^*$ into the main diagonal, 2nd, 4th, 6th and etc. sub-diagonal block
positions. This kind of $\bW$ corresponds to the pooling of $\sqrt{p}$
districts (each district has $\sqrt{p}$ locations) which the evenly
numbered districts are connected and the oddly numbered districts are
connected but evenly numbered districts and oddly number districts are
separated. Each district has similar neighboring structures. As $p$
increases, the number of the locations influencing one specific location
increases in the order of $\sqrt{p}$. The error $\varepsilon_{i,t}$ are
independently generated from $N(0,\sigma_i^2)$, where we generate each
$\sigma_i$ from $U(0.5,1.5)$.

\begin{figure}[ht]
	\centering
	\mbox{ \resizebox{11cm}{!}{\includegraphics{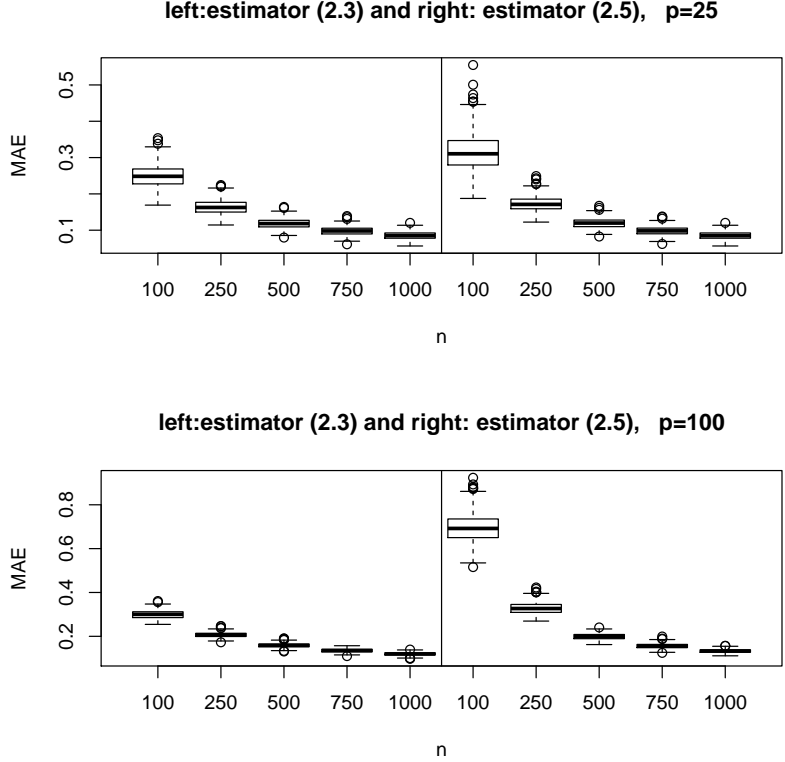}} }
\caption{\label{figure3} Boxplots of MAE for estimator (2.3) (left panels) and
estimator (2.5) (right panels) with $p=25 $ (top panels) and 100 (bottom panels), $n=100,\,250,\,500,\,750,\,1000$ for scenario 2.}
\end{figure}

Figure \ref{figure3} depicts two boxplots of MAE with $p$ equals to, respectively, 25 and 100. As the sample size $n$ increases from 100, 250, 500, 750 to 1000, MAE decreases for both methods.

\begin{figure}[ht]
	\centering
	\mbox{ \resizebox{11cm}{!}{\includegraphics{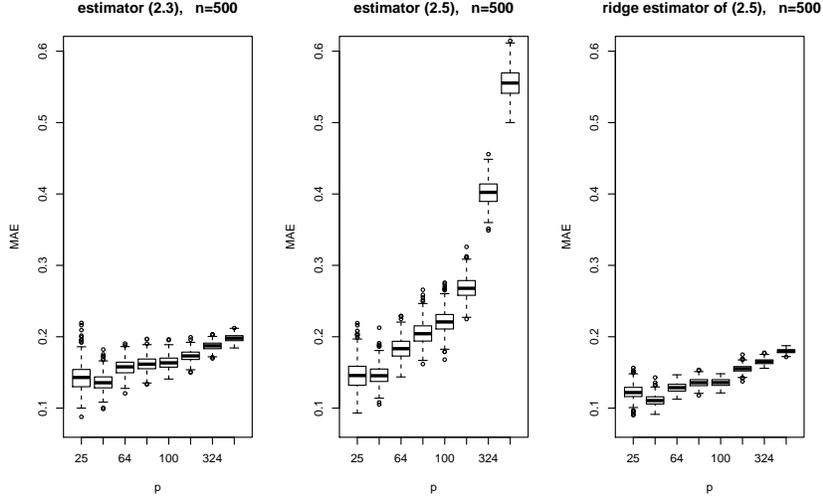}} }
\caption{\label{figure4} Boxplots of MAE of the original estimator (2.3) (the left panel),
the root $n$ consistent estimator (2.5) (the middle panel), and the estimator (2.5) after adding ridge penalty (the right panel)  with $n=500$ and $p=25,49,64,81,100,169,324,529$ for scenario 2.}
\end{figure}

Figure \ref{figure4} depicts three boxplots as Figure
\ref{figure2}. The MAE for (2.3) increases steadily as $p$ increases,
which matches the result of Theorem 4 when, for instance, $s_1(p) \asymp
\sqrt{p}$, $s_0(p) \asymp p $ and $ s_2(p) \asymp p$.
The MAE for (2.5) after adding ridge penalty is slowly increasing as well. This
might be caused by the fact that, similar to A2(c), quantities in
condition A4(a) is also influenced by $p$ since the number of nonzero
elements in $\bw_i$ is in the order of $\sqrt{p}$.

\section{Real data analysis} \label{real data}

\subsection{European Consumer Price Indices}

We analyze the monthly change rates of the consumer price index (CPI) for
the EU member states, over the years 2003-2010. We use the national
harmonized index of consumer prices calculated by Eurostat, the
statistical office of the European Union.
For this data set, $n=96$ and $p=31$.

\begin{figure}[ht]
	\centering
\mbox{ \resizebox{13cm}{4cm}{\includegraphics{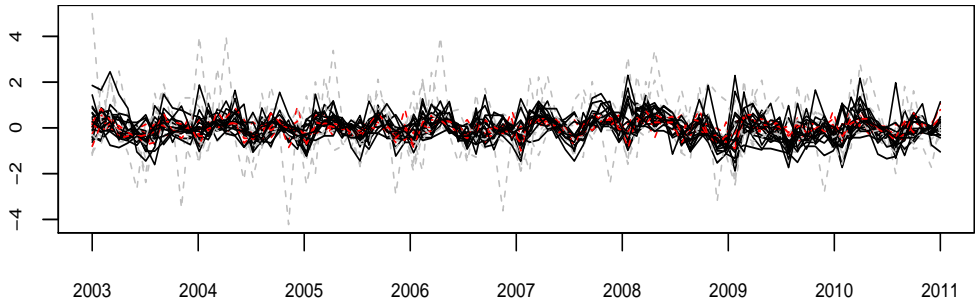}} }
\caption{\label{plot1} \small Time series plots of the monthly change
rates of CPI for the 31 EU member states.
Each series is subtracted by its mean value.}
\end{figure}

Figure \ref{plot1} presents the time series plots of the monthly change
rates of CPI for the 31 states. To line up the curves together,
 each series is centered at its mean value in Figure \ref{plot1}.
There exist clearly synchronizes on the fluctuations across different
states, indicating the spatial (i.e. cross-state) correlations
among different states. Also noticeable is the varying degrees of
the fluctuation over the different states.

Let $\by_t$ consist of the monthly change rates of CPI for the 31 states.
We fit the proposed spatial-temporal model (\ref{fullmodel}) to this data
set with the parameters estimated by (\ref{estimator}).
We take a normalized sample correlation matrix  of
$\by_t$ as the spatial weight matrix $\bW = (w_{ij})$, i.e. we let
$w_{ij}$ be the absolute value of the sample correlation between the $i$-th and $j$-th states
for $i\ne j$, and $w_{ii}=0$, and then replace $w_{ij}$ by $w_{ij}/\sum_k w_{kj}$.

\begin{figure}[ht]
	\centering
        \mbox{ \resizebox{3.6cm}{!}{\includegraphics{focus_belgium0.eps}} }
        \mbox{ \resizebox{3.6cm}{!}{\includegraphics{focus_greece0.eps}} }
        \mbox{ \resizebox{3.6cm}{!}{\includegraphics{focus_france0.eps}} }
        \mbox{ \resizebox{3.6cm}{!}{\includegraphics{focus_iceland0.eps}} }
        \mbox{ \resizebox{3.6cm}{!}{\includegraphics{focus_belgium1.eps}} }
        \mbox{ \resizebox{3.6cm}{!}{\includegraphics{focus_greece1.eps}} }
        \mbox{ \resizebox{3.6cm}{!}{\includegraphics{focus_france1.eps}} }
        \mbox{ \resizebox{3.6cm}{!}{\includegraphics{focus_iceland1.eps}} }
        \mbox{ \resizebox{3.6cm}{!}{\includegraphics{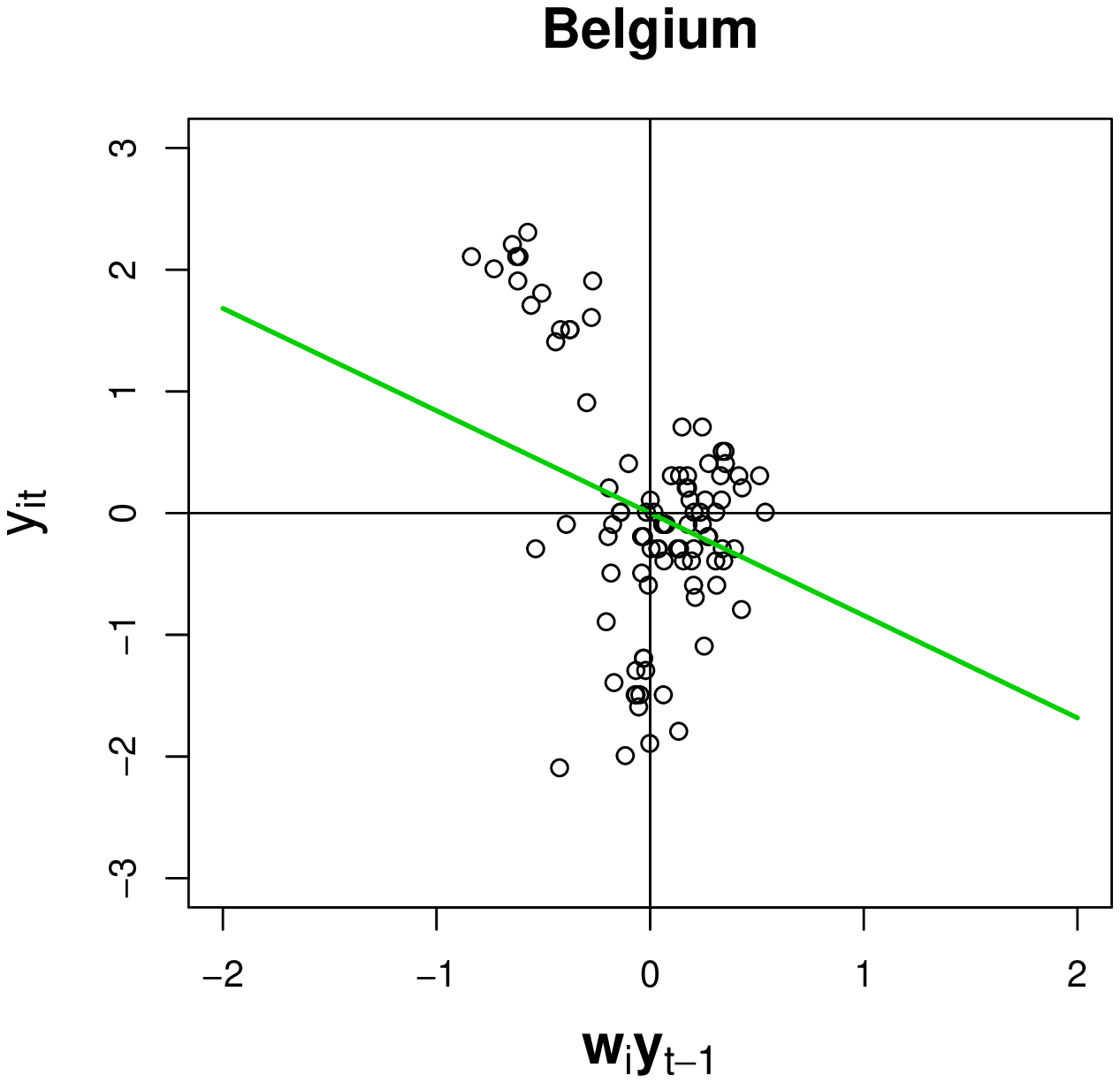}} }
        \mbox{ \resizebox{3.6cm}{!}{\includegraphics{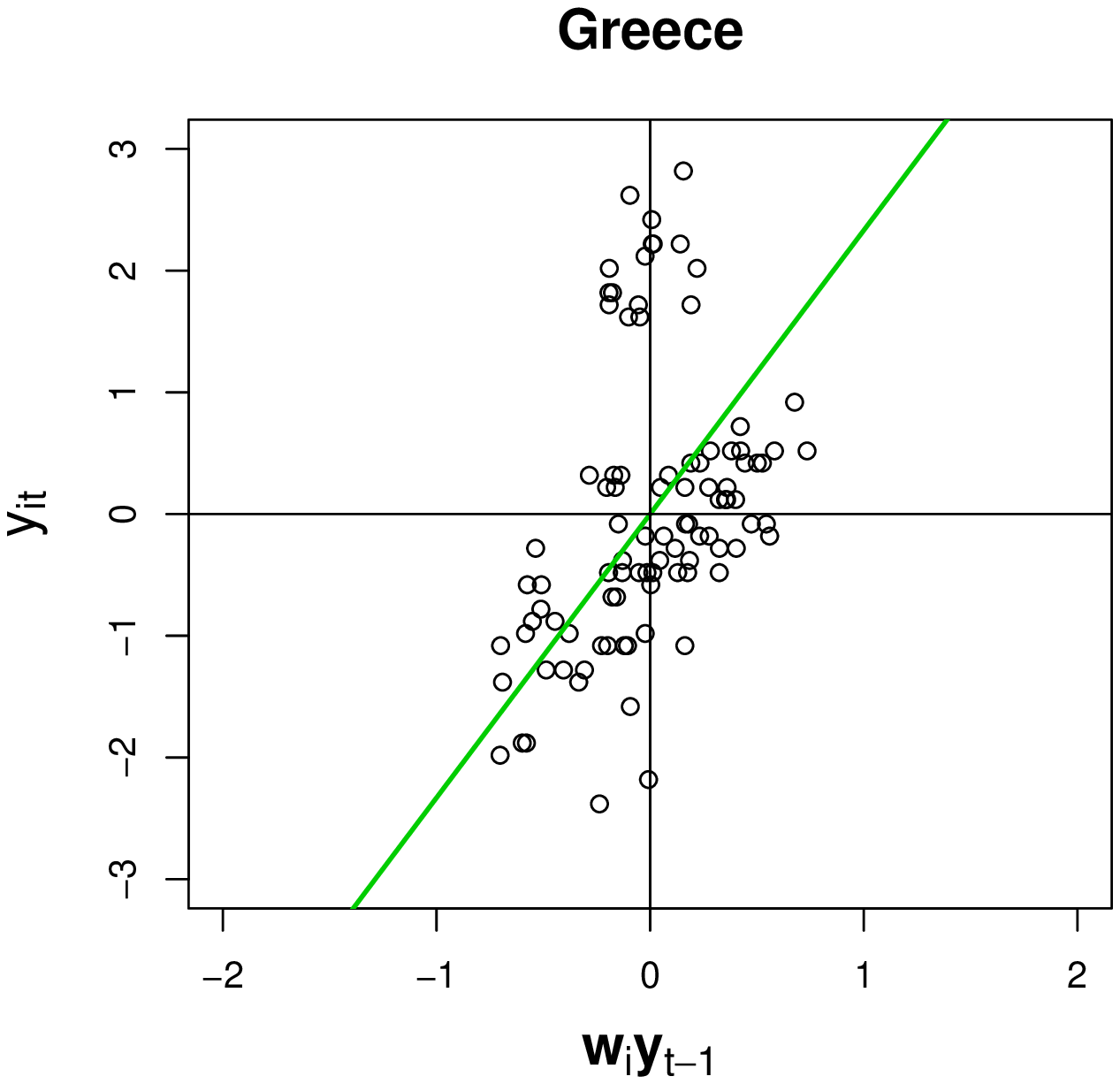}} }
        \mbox{ \resizebox{3.6cm}{!}{\includegraphics{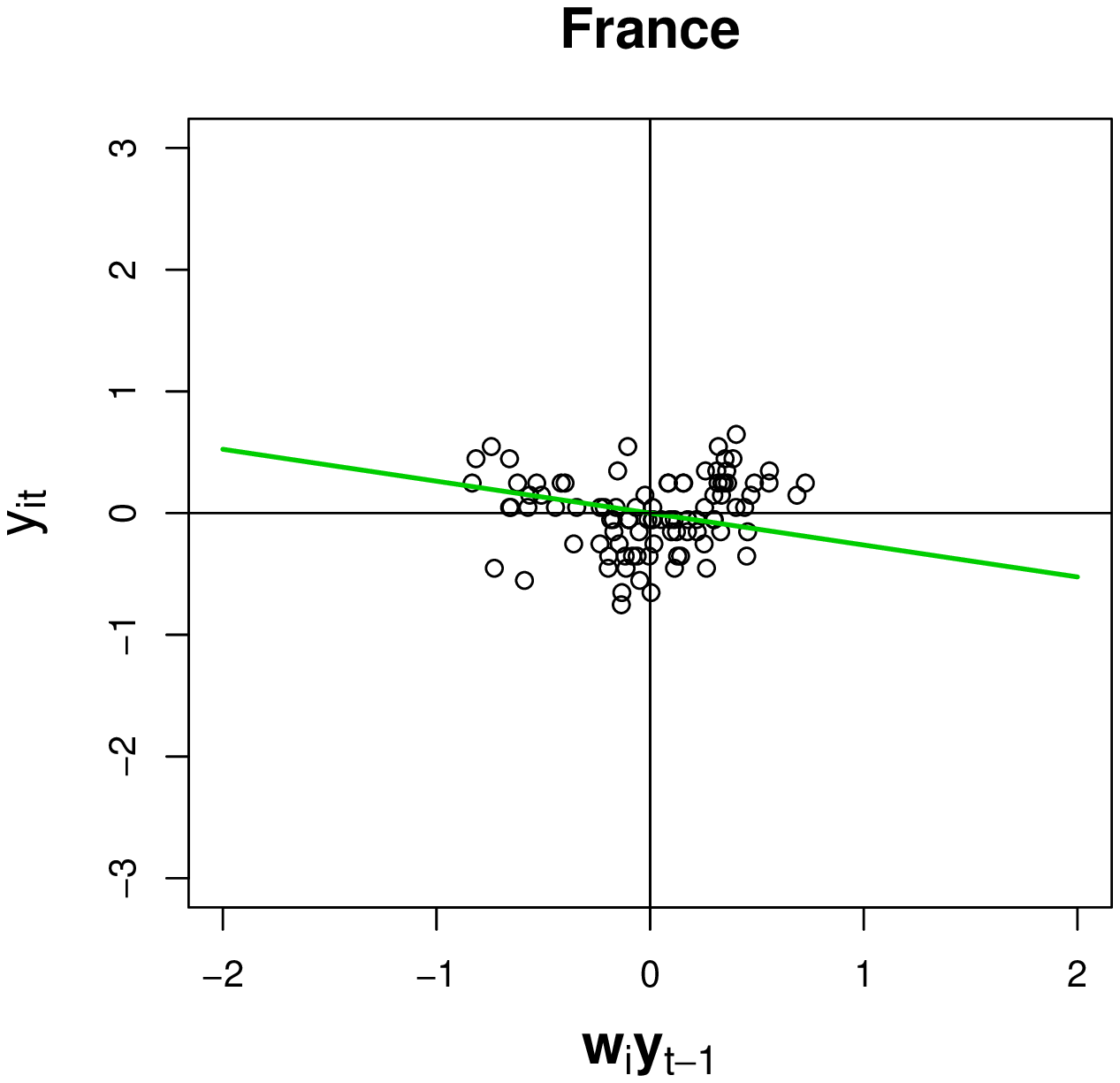}} }
        \mbox{ \resizebox{3.6cm}{!}{\includegraphics{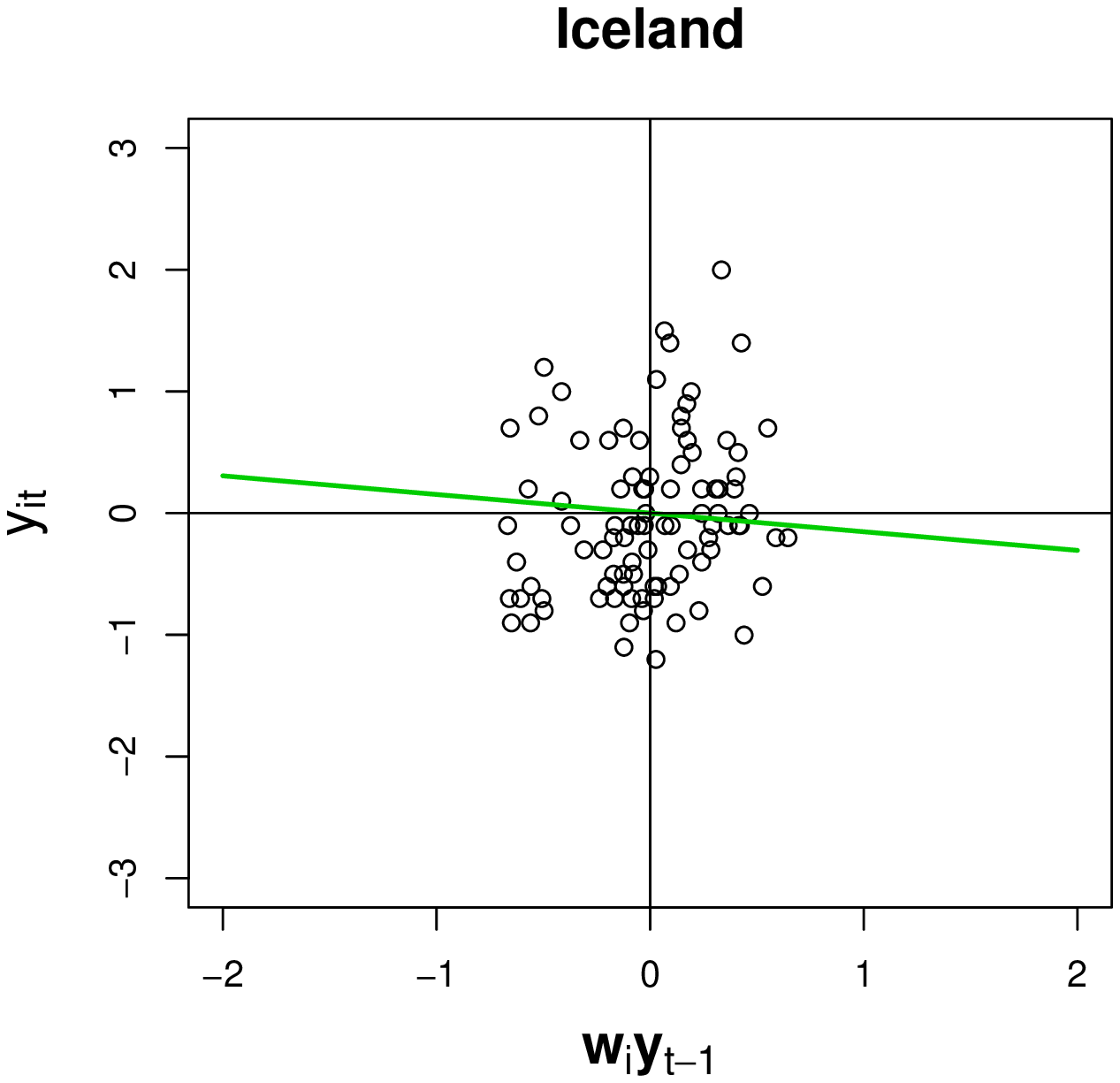}} }
\caption{\label{fig_appl1a}
The scatter plots of $y_{i,t}$ against $\bw^T_i \by_t$ (panels on the top),
$y_{i, t-1}$ (panels in the middle), and $\bw^T_i \by_{t-1}$ (panels on
the bottom) for four selected countries
Belgium, Greece, France and Iceland. The straight lines $y=\wh \lambda_{ji} x $
are superimposed in the panels on the top with $j=0$, those in the
middle with $j=1$, and those on the bottom with $j=2$.}
\end{figure}

\begin{figure}[ht]
	\centering
	\mbox{ \resizebox{12cm}{!}{\includegraphics{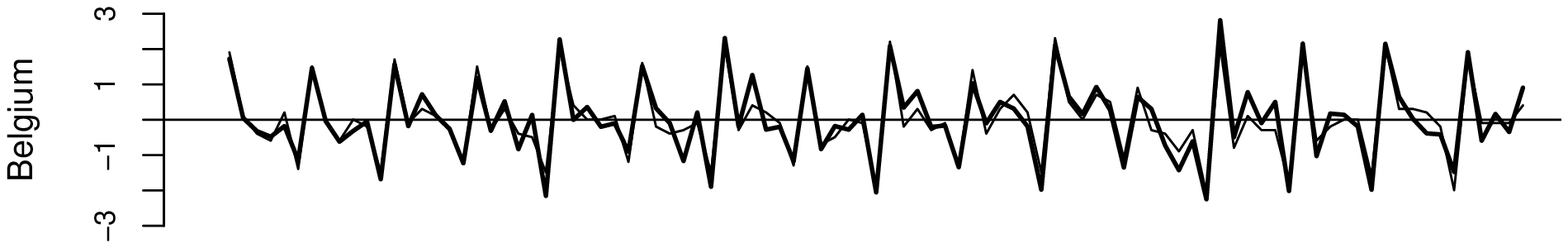}} }
	\mbox{ \resizebox{12cm}{!}{\includegraphics{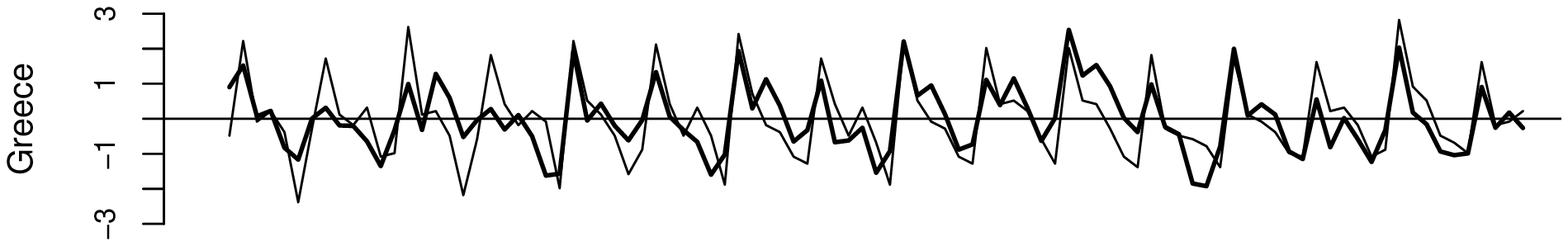}} }
	\mbox{ \resizebox{12cm}{!}{\includegraphics{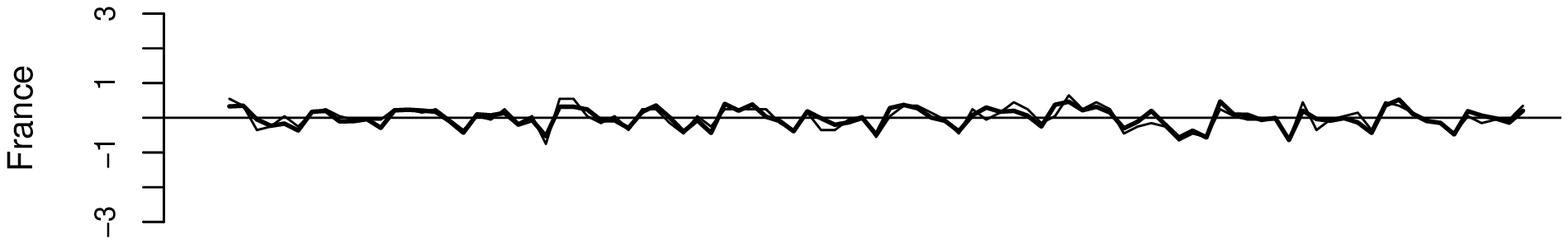}} }
    \mbox{ \resizebox{12cm}{!}{\includegraphics{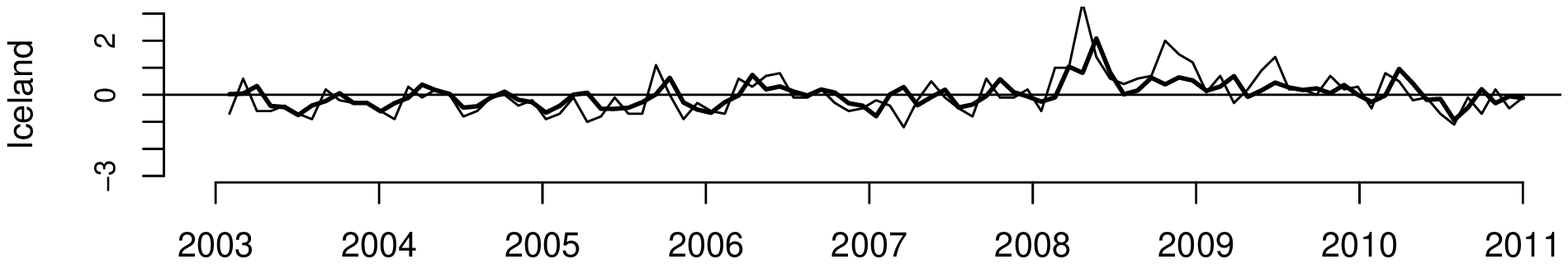}} }
\caption{\label{t1-plots} \small The monthly change rates  of
CPI (thin lines) of Belgium, Greece, France and Iceland, and their estimated values (thick lines) by
model (\ref{fullmodel}).}
\end{figure}

Figure~\ref{fig_appl1a} presents the scatter plots of $y_{i,t}$ against, respectively,
the 3 regressors in model (\ref{fullmodel}), i.e. $\bw_i^T\by_t, \,
y_{i, t-1}, \, \bw_i^T\by_{t-1}$, for four selected states Belgium, Greece, France and
Iceland. We superimpose the straight line $y= \wh \lambda_{ji}\, x$ in each of those
3 scatter plots with, respectively, $j=0, 1, 2$.
It is clear that the estimated slopes are very different for those 4 states.
Figure~\ref{t1-plots} plots the true monthly change rates
of the CPI for those 4 states together with the fitted values
\begin{equation} \label{resid}
\wh y_{i,t} = \wh \lambda_{0i} \bw_i^T\by_t + \wh \lambda_{1i} y_{i, t-1}
+ \wh \lambda_{2i} \bw_i^T\by_{t-1}. \tag{12}
\end{equation}

\begin{figure}[ht]
	\centering
	\mbox{ \resizebox{!}{6cm}{\includegraphics{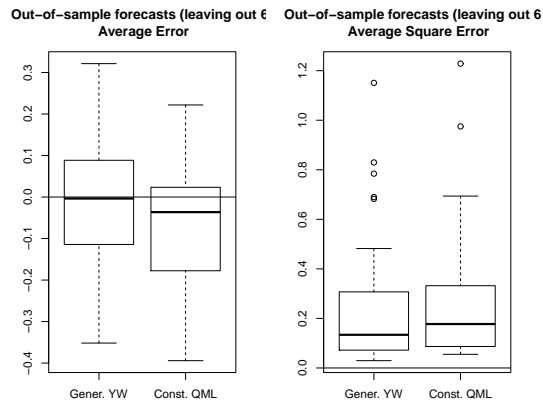}} }
\caption{\label{out_sample} \small Prediction errors generated in the out-of-sample forecasting, leaving out 6 observations from the sample, using our model with the Generalized Yule-Walker estimator and using the constant SDPD model of Yu et al. (2008) with the Quasi-Maximum Likelihood estimator.}
\end{figure}

Overall $\wh y_{i,t}$ tracks its truth value reasonably well.
Figure \ref{out_sample} shows the out-of-sample forecasting
performance of our model. For the sake of comparison, predictions are
made using our model and the proposed generalized Yule-Walker
estimator, and using the (constant) SDPD model of Yu et al. (2008) and their
Quasi-Maximum Likelihood estimator. In particular, for each location, we
leave out from the sample the last six observations and we compute the
(out-of-sample) forecasts with 1,2,....6 step ahead forecasting horizon;
then, we compute the average prediction error over time (i.e. the mean of the 6
prediction errors).  On the left panel of Figure \ref{out_sample}, the
two box-plots summarize the average prediction error for the 31 locations
obtained with our YW estimator and the QML estimator of Yu et al. (2008),
respectively. It is evident that our estimator produces unbiased
predictions while the QML estimator appears to be
biased.
This advantage also reflects on the forecasting average square errors,
reported on the right panel of Figure \ref{out_sample}. In conclusion,
the SDPD model of Yu et al. (2008) has a satisfying forecasting performance
because several locations have similar spatial structure and for those
locations a model with constant parameters is sufficient. Anyway, a
marginal improvement is observed for our estimator because several
locations have quite different structures and  our model is able to
capture this difference. Finally,  it is worthwhile to notice that the
variability of the two predictors appears to be the same. 


To further vindicate the necessity to use different coefficients for different
states, we consider a statistical test for hypothesis
\[
H_0: \lambda_{j1} = \cdots  = \lambda_{jp}, \;\; j=0, \, 1, \, 2
\]
for model (\ref{fullmodel}).
Then the residuals resulting from the fitted model under $H_0$ will be greater than
the residuals without $H_0$. However if $H_0$ is true, the difference between the
two sets of residuals should not be significant. We apply a bootstrap method
to test this significance.
Let $\wt \lambda_0, \wt \lambda_1, \wt \lambda_2$
be the estimates under hypothesis $H_0$. Define the test statistic
\begin{equation} \label{test}
U= {1 \over n}  \sum_{t=1}^n \|\by_t - \wt \by_t\|_1, \qquad
\wt \by_t = \wt \lambda_0 \bW \by_t + \wt \lambda_1 \by_{t-1} + \wt \lambda_2 \bW \by_{t-1}. \nonumber
\end{equation}
We reject $H_0$ for large values of $U$. To assess how large is large, we generate
a bootstrap data from
\begin{equation} \label{boots2}
\by_t^* = \wt \lambda_0 \bW \by_t + \wt \lambda_1 \by_{t-1} + \wt \lambda_2 \bW \by_{t-1}
+ \bve_t^*, \nonumber
\end{equation}
where $\{ \bve_t^*\}$ are drawn independently from the residuals
\[
\wh \bve_t = \by_t - \wh\by_t, \qquad t=1, \cdots, n,
\]
and $\wh \by_t$ consists of
the components defined in (\ref{resid}).
Now the bootstrap statistic is defined as
\[
U^* = {1 \over n} \sum_{t=1}^n \|\by^*_t -(\lambda_0^* \bW \by_t +
\lambda_1^* \by_{t-1} +  \lambda_2^* \bW \by_{t-1})\|_1 ,
\]
where $(\lambda_0^*, \lambda_1^*, \lambda_2^*)$ is the estimated coefficients for the
regression model
\[
\by_t^* =  \lambda_0 \bW \by_t +  \lambda_1 \by_{t-1} +  \lambda_2 \bW \by_{t-1}
+ \bve_t, \qquad t=1, \cdots, n.
\]
The $P$-value
for testing hypothesis $H_0$ is defined as
\[
P( U^*>U|\by_1, \cdots, \by_n),
\]
which is approximated by the relative frequency of the event $(U^*>U)$ in a
repeated bootstrap sampling with a large number of replications.
By repeating bootstrap sampling 1000 times, the estimated $P$-value is 0, exhibiting
strong evidence against the null hypothesis $H_0$. Therefore the model with
the equal slope parameters across different locations is inadequate for this
particular data set.

\subsection{Modeling mortality rates}

\begin{figure}[t]
	\centering
	\mbox{ \resizebox{12cm}{!}{\includegraphics{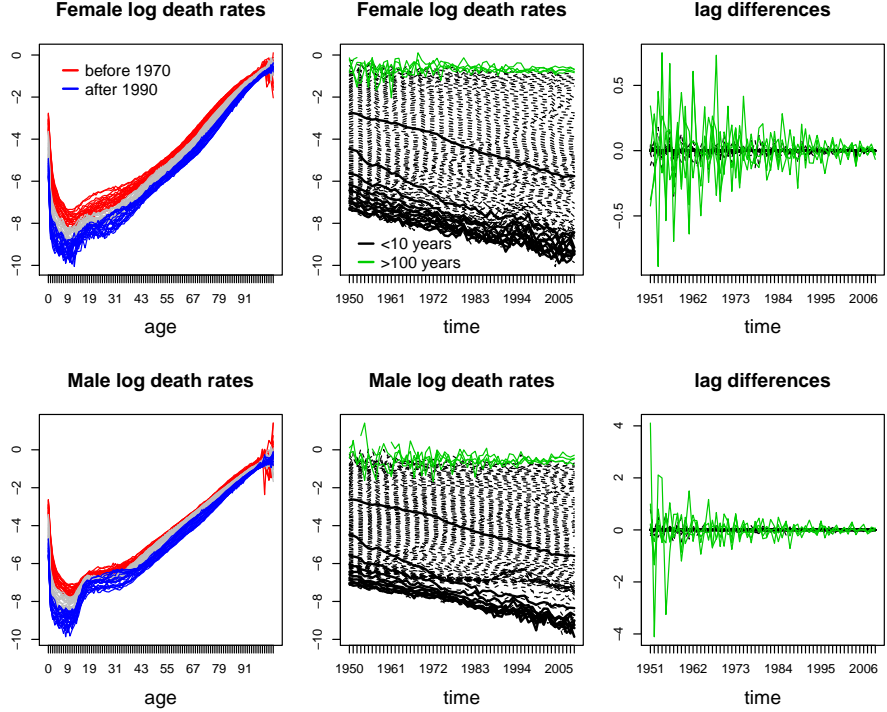}} }
\caption{\label{fig_appl1} \small Log mortality rates of Italian female
(3 top panels) and male (3 bottom panels) are plotted against age from
each year in 1950-2009 (2 left panels), against year for each age group between 0 and
104  (2 middle panels). Differenced log mortality
rates are plotted against year for each age in 2 right panels.}
\end{figure}

Now we analyze the annual Italian male and female mortality
rates for different ages (between 0 and 104)  in the period of 1950 -- 2009
based on the proposed model (\ref{fullmodel}).
The data were downloaded from the Human Mortality Database (see the
website http://www.mortality.org/).
Let $m_{i,t}$ be  the log mortality rate of female or male
at age $i$ and in Year $t$. Those data are plotted in Figure \ref{fig_appl1}.
Two panels on the left plot are the female and male mortality
against different age in each year. More precisely the curves $\{
m_{i, t}, \, i=1, \cdots, 21\}$ for $t< 1970$ are plotted in red,
those for $t >1990$ are in blue, those with $1970\le t \le 1989$ are in grey.
Those curves show clearly that the mortality rate decreases over the years for
almost all age groups (except a few outliers at the top end).
Two panels in the middle of Figure \ref{fig_appl1} plot the log mortality
for each age group
against time
with the following color code:
black for ages not great than  10,
grey for ages between 11 and 100, and green for ages greater than 100.
They indicate that the mortality for all age groups decreases over time,
the most significant decreases occur at the young age groups.
Furthermore, the fluctuation of
the mortality rates for the top age groups reduces significantly over the years,
while the mean mortality rates for those groups remain about the same. This
can be seen more clearly in the two panels on the right 
which plot differenced log mortality rates
$\{ y_{i, t}, \, t=1951, \cdots, 2009\}$,
using the same colour code,
where
$
y_{i, t} = m_{i, t} - m_{i, t-1}.
$

We fit the differenced log mortality data with model (\ref{fullmodel}) with the
parameters estimated by (\ref{new_estimator}) and $d_i=20$. Note that now $p=104$ and
$n=59$. Let
the off-diagonal elements of the spatial weight matrix $\bW$ be
\[
w_{ij} = {1 \over 1 +|i-j|}, \qquad 1\le i < j \le 104.
\]
We then replace $w_{ij}$ by $w_{ij}/ \sum_i w_{ij}$. Moreover, we can also fix a threshold $\tau$ and set to zero all the elements of matrix $\bW$ such that $|x-w|>\tau$ (for simplicity, we fix $\tau=5$ in this application, but the results are substantially invariant for different values of $\tau$).

\begin{figure}[ht]
	\centering
	\mbox{ \resizebox{12cm}{!}{\includegraphics{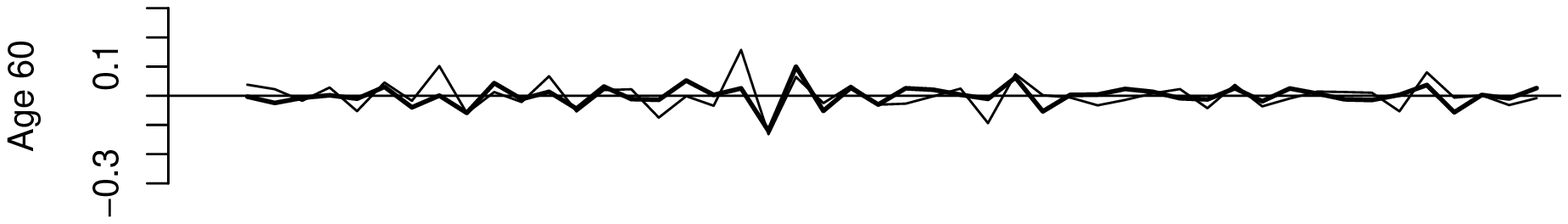}} }
	\mbox{ \resizebox{12cm}{!}{\includegraphics{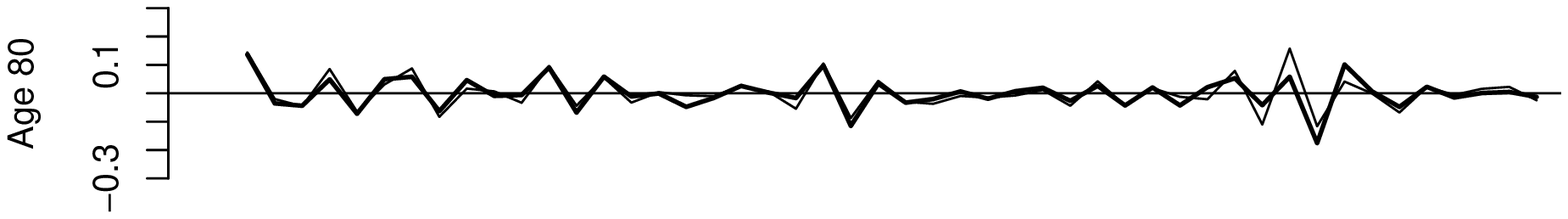}} }
	\mbox{ \resizebox{12cm}{!}{\includegraphics{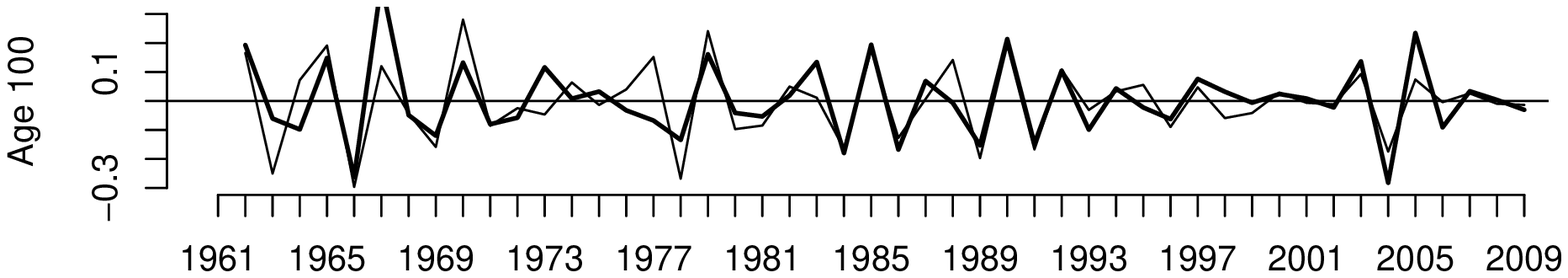}} }
\caption{\label{t-plots} \small Observed time series (thin line) and fitted time series (bold line), for female mortality rate for ages $i=60,80,100$.}
\end{figure}

The results of the estimation are shown in table \ref{table_values}, for a selection of cohorts of different ages.
Figure \ref{t-plots} shows the fitted series for ages $i=60,80,100$.
\begin{table}[t]
\centering
\begin{tabular}{|c|ccc|c|ccc|} \hline
age & $\hat\lambda_{0i}$ & $\hat\lambda_{1i}$ & $\hat\lambda_{2i}$ & age & $\hat\lambda_{0i}$ & $\hat\lambda_{1i}$ & $\hat\lambda_{2i}$ \\ \hline
5  &  0.41 & -0.52 &  0.06 & 55 &  0.19 & -0.88 &  0.28 \\
10 &  0.20 & -0.42 &  0.05 & 60 & -0.09 & -0.72 &  0.01 \\
15 &  0.44 & -0.65 &  0.18 & 65 &  0.22 & -0.63 &  0.21 \\
20 &  0.64 & -0.78 &  0.40 & 70 &  0.21 & -0.69 &  0.08 \\
25 & -0.04 & -0.43 &  0.03 & 75 &  0.33 & -0.59 &  0.22 \\
30 &  0.78 & -0.80 &  0.55 & 80 &  0.33 & -0.89 &  0.27 \\
35 &  0.11 & -0.55  & 0.29 & 85 &  0.37 & -0.76 &  0.18 \\
40 & -0.04 & -0.66 & -0.01 & 90 &  0.29 & -0.62 &  0.16 \\
45 &  0.29 & -0.46 &  0.12 & 95 &  0.27 & -0.77 &  0.26 \\
50 & -0.10 & -0.45 & -0.05 & 100 &  0.44 & -0.69 & -0.03 \\ \hline
\end{tabular}
\caption{\label{table_values} \small Estimated coefficients for a selection of cohorts of different ages. The left column is the estimated pure spatial coefficients $\hat\lambda_{0i}$; The middle column is the estimated pure dynamic coefficient $\hat\lambda_{1i}$; The right column is the estimated spatial-dynamic coefficients $\hat\lambda_{2i}$.}
\end{table}

\section{Final remark}

We propose in this paper a generalized Yule-Walker estimation method for spatio-temporal
models with  diagonal coefficients. The setting enlarges the capacity of the
popular spatial dynamic panel data models. Both the asymptotic results and numerical
illustration show that the proposed estimation method works well,
although the number of the estimation equations utilized should be of the order $o(\sqrt{n})$.

\section*{Appendix: Proofs}
\renewcommand{\theequation}{7.\arabic{equation}}
\label{appendix}

We present the proofs for Theorems 2, Corollary 1 and Theorem 4 in this appendix.
The proofs for Theorem 1 and 3 are similar and simpler than that of Theorem 2, and they are
therefore omitted.
We also present a lemma (i.e. Lemma 1) at the end of this appendix, which
shows that condition A2 is implied by conditions
A1 and B1 -- B3; see Remark 1.
We use $C$ to denote a generic positive constant, which may be different at different places.

\askip

\noindent
{\bf Proof of Theorem 2.}
We first prove (i) of Theorem 2.
We only need to prove the assertions (1) and (2) below, as then the required
conclusion follows from (1) and (2) immediately.
\begin{itemize}
\item[(1)]
\begin{equation}
\sqrt{n} \bU_i^{-\frac{1}{2}}  \left(
\begin{array}{c}
\frac{1}{n}\sum_{t=1}^{n}\by_{t-1}^T(\bw_i^T\by_t) \frac{1}{n}\sum_{t=1}^{n}\varepsilon_{i, t}\by_{t-1}  \\
\frac{1}{n}\sum_{t=1}^{n}\by_{t-1}^Ty_{i,t-1} \frac{1}{n}\sum_{t=1}^{n}\varepsilon_{i, t}\by_{t-1} \\
\frac{1}{n}\sum_{t=1}^{n}\by_{t-1}^T(\bw_i^T\by_{t-1}) \frac{1}{n}\sum_{t=1}^{n}\varepsilon_{i, t}\by_{t-1}
\end{array}
\right)
\xrightarrow{d} N(0,\bI_3). \nonumber
\end{equation}
\item[(2)]
$
\bV_i (\hat \bX_i^T\hat \bX_i)^{-1} \xrightarrow{P} \bI_3. \nonumber
$
\end{itemize}

To prove (1), it suffices to show that for any nonzero vector $\ba=(a_1,a_2,a_3)^{T}$, the linear combination
\[
\ba^{T}\left(
\begin{array}{c}
\frac{1}{n}\sum_{t=1}^{n}\by_{t-1}^T(\bw_i^T\by_t) \frac{1}{n}\sum_{t=1}^{n}\varepsilon_{i, t}\by_{t-1}  \\
\frac{1}{n}\sum_{t=1}^{n}\by_{t-1}^Ty_{i,t-1} \frac{1}{n}\sum_{t=1}^{n}\varepsilon_{i, t}\by_{t-1} \\
\frac{1}{n}\sum_{t=1}^{n}\by_{t-1}^T(\bw_i^T\by_{t-1}) \frac{1}{n}\sum_{t=1}^{n}\varepsilon_{i, t}\by_{t-1}
\end{array}
\right)
\]
is asymptotic normal.

Let us take out the dominant term in $\frac{1}{n}\sum_{t=1}^n \by_{t-1}^{T}(\bw_i^T\by_t)\frac{1}{n}\sum_{t=1}^n \varepsilon_{i,t}\by_{t-1}$ first.
\begin{equation} \label{error_decom}
\begin{split}
& \frac{1}{n}\sum_{t=1}^n \by_{t-1}^{T}(\bw_i^T\by_t)\frac{1}{n}\sum_{t=1}^n \varepsilon_{i,t}\by_{t-1}
\\ = & \left[\frac{1}{n}\sum_{t=1}^n \by_{t-1}^{T}(\bw_i^T\by_t)-\mathrm{E}[\by_{t-1}^{T}(\bw_i^T\by_t)]\right]\frac{1}{n}\sum_{t=1}^n \varepsilon_{i,t}\by_{t-1}+
\mathrm{E}[\by_{t-1}^{T}(\bw_i^T\by_t)]\frac{1}{n}\sum_{t=1}^n \varepsilon_{i,t}\by_{t-1}
\\ = & \left[\frac{1}{n}\sum_{t=1}^n \by_{t-1}^{T}(\bw_i^T\by_t)-\bw_i^T\bSigma_1\right]\frac{1}{n}\sum_{t=1}^n \varepsilon_{i,t}\by_{t-1}+\frac{1}{n}\sum_{t=1}^n \bw_i^T\bSigma_1\by_{t-1}\varepsilon_{i,t}
\\ = & E_1+E_2.
\end{split} \tag{13}
\end{equation}
For term $E_1$ and $k=1,2,\cdots,p$, by Proposition 2.5 of Fan and Yao (2003), we have
\begin{equation} \label{mixing_bound}
\begin{split}
& \mathrm{E}\left[\frac{1}{n}\sum_{t=1}^n (\be_k^T\by_{t-1}\bw_i^T\by_t-\be_k^T\bSigma_1^T\bw_i)\right]^2
\\ = & \frac{1}{n^2}\sum_{t=1}^{n}\mathrm{Var}(\be_k^T\by_{t-1}\bw_i^T\by_t)+\frac{1}{n^2}\sum_{t \neq s}\mathrm{Cov}(\be_k^T\by_{t-1}\bw_i^T\by_t, \be_k^T\by_{s-1}\bw_i^T\by_s)
\\ \le & \frac{C}{n}+\frac{1}{n^2}\sum_{t \neq s}8\alpha(|t-s|)^{\frac{\gamma}{4+\gamma}}\left[\mathrm{E}|\be_k^T\by_{t-1}\bw_i^T\by_t|^{2+\frac{\gamma}{2}}\right]^{\frac{2}{4+\gamma}}
\left[\mathrm{E}|\be_k^T\by_{s-1}\bw_i^T\by_s|^{2+\frac{\gamma}{2}}\right]^{\frac{2}{4+\gamma}}
\\ \le & \frac{C}{n}+\frac{C}{n^2}\sum_{t \neq s}\alpha(|t-s|)^{\frac{\gamma}{4+\gamma}} \le  \frac{C}{n}+\frac{C}{n} \sum_{j=1}^{n}\alpha(j)^{\frac{\gamma}{4+\gamma}}=O(\frac{1}{n}),
\end{split} \tag{14}
\end{equation}
where $C$ is independent of $p$. Then it holds that
\begin{equation}
\frac{1}{n}\sum_{t=1}^n (\be_k^T\by_{t-1}\bw_i^T\by_t-\be_k^T\bSigma_1^T\bw_i)=O_p(\frac{1}{\sqrt{n}}). \nonumber
\end{equation}
Therefore
\begin{equation}
\left \| \frac{1}{n}\sum_{t=1}^n \by_{t-1}\bw_i^T\by_t-\bSigma_1^T\bw_i \right \|_2=\sqrt{\sum_{k=1}^p\left[\frac{1}{n}\sum_{t=1}^n (\be_k^T\by_{t-1}\bw_i^T\by_t-\be_k^T\bSigma_1^T\bw_i)\right]^2}=O_p(\sqrt{\frac{p}{n}}). \nonumber
\end{equation}
Similarly,
\begin{equation}
\left\|\frac{1}{n}\sum_{t=1}^n \varepsilon_{i,t}\by_{t-1}\right\|_2=O_p(\sqrt{\frac{p}{n}}). \nonumber
\end{equation}
Since $E_1 \le \left \| \frac{1}{n}\sum_{t=1}^n \by_{t-1}\bw_i^T\by_t-\bSigma_1^T\bw_i \right \|_2\left\|\frac{1}{n}\sum_{t=1}^n \varepsilon_{i,t}\by_{t-1}\right\|_2$, it holds that $E_1=O_p(\frac{p}{n})$. Similar to (\ref{mixing_bound}), we have $\mathrm{Var}(\sqrt{n}E_2)=O(1)$. Given $\frac{p}{\sqrt{n}}=o(1)$, it holds that $\sqrt{n}E_1=o_p(1)$. Hence if $p=o(\sqrt{n})$,
\begin{equation}
\begin{split}
& \sqrt{n} \times \frac{1}{n}\sum_{t=1}^n \by_{t-1}^{T}(\bw_i^T\by_t)\frac{1}{n}\sum_{t=1}^n \varepsilon_{i,t}\by_{t-1}=\frac{1}{\sqrt{n}}\sum_{t=1}^n \bw_i^T\bSigma_1\by_{t-1}\varepsilon_{i,t}+o_p(1).
\end{split} \nonumber
\end{equation}
Similarly, given $p=o(\sqrt{n})$, we have
\begin{equation}
\begin{split}
& \sqrt{n} \times \frac{1}{n}\sum_{t=1}^{n}\by_{t-1}^Ty_{i,t-1}  \frac{1}{n}\sum_{t=1}^{n}\varepsilon_{i, t}\by_{t-1}=\frac{1}{\sqrt{n}}\sum_{t=1}^n \be_i^T\bSigma_0\by_{t-1}\varepsilon_{i,t}+o_p(1),
\end{split} \nonumber
\end{equation}
\begin{equation}
\begin{split}
& \sqrt{n} \times \frac{1}{n}\sum_{t=1}^{n}\by_{t-1}^T(\bw_i^T\by_{t-1})  \frac{1}{n}\sum_{t=1}^{n}\varepsilon_{i, t}\by_{t-1}=\frac{1}{\sqrt{n}}\sum_{t=1}^n \bw_i^T\bSigma_0\by_{t-1}\varepsilon_{i,t}+o_p(1).
\end{split} \nonumber
\end{equation}
Now it suffices to prove
\begin{equation}
S_{n,p} \equiv a_1\frac{1}{\sqrt{n}}\sum_{t=1}^n \bw_i^T\bSigma_1\by_{t-1}\varepsilon_{i,t}+a_2\frac{1}{\sqrt{n}}\sum_{t=1}^n \be_i^T\bSigma_0\by_{t-1}\varepsilon_{i,t}+a_3\frac{1}{\sqrt{n}}\sum_{t=1}^n \bw_i^T\bSigma_0\by_{t-1}\varepsilon_{i,t} \nonumber
\end{equation}
is asymptotic normal.

Note that it holds that
\[
\mathrm{E}|\bw_i^T\bSigma_1\by_{t-1}\varepsilon_{i,t}|^{2+\frac{\gamma}{2}} \le [\mathrm{E}|\bw_i^T\bSigma_1\by_{t-1}|^{4+\gamma}]^{\frac{1}{2}}
[\mathrm{E}|\varepsilon_{i,t}|^{4+\gamma}]^{\frac{1}{2}}<\infty.
\]
Now we calculate the variance of $S_{n,p}$. It holds that
\begin{equation} \label{variance}
\begin{split}
& \var\left(\frac{1}{\sqrt{n}}\sum_{t=1}^n \bw_i^T\bSigma_1\by_{t-1}\varepsilon_{i,t}\right)
\\ = & \bw_i^T\bSigma_1\bSigma_{\by, \varepsilon_{i}}(0) \bSigma_1^{T}\bw_i+\sum_{j=1}^{n-1}\left( 1-\frac{j}{n} \right)\bw_i^T\bSigma_1\left[\bSigma_{\by, \varepsilon_{i}}(j)+\bSigma_{\by, \varepsilon_{i}}^T(j)\right] \bSigma_1^{T}\bw_i,
\end{split} \tag{15}
\end{equation}
and it follows from $\sum_{j=1}^{n}\alpha(j)^{\frac{\gamma}{4+\gamma}}< \infty$ that
\begin{equation*}
\begin{split}
& \sup_{p}\sum_{j=1}^{\infty}|\bw_i^T\bSigma_1\left[\bSigma_{\by, \varepsilon_{i}}(j)+\bSigma_{\by, \varepsilon_{i}}^T(j)\right] \bSigma_1^{T}\bw_i|
\\ \le & C \sup_{p}\sum_{j=1}^{\infty}\alpha(j)^{\frac{\gamma}{4+\gamma}} \left\{\mathrm{E}|\bw_i^T\bSigma_1\by_{t-1}
|^{4+\gamma}\right\}^{\frac{2}{4+\gamma}} \left\{\mathrm{E}|\varepsilon_{i,t}
|^{4+\gamma}\right\}^{\frac{2}{4+\gamma}}< \infty.
\end{split}
\end{equation*}
Similarly,
\begin{equation}
\begin{split}
& \mathrm{Cov}\left(\frac{1}{\sqrt{n}}\sum_{t=1}^n \bw_i^T\bSigma_1\by_{t-1}\varepsilon_{i,t},\frac{1}{\sqrt{n}}\sum_{t=1}^n \be_i^T\bSigma_0\by_{t-1}\varepsilon_{i,t} \right)
\\ = & \bw_i^T\bSigma_1\bSigma_{\by, \varepsilon_{i}}(0) \bSigma_0^{T}\be_i+\sum_{j=1}^{n-1}\left( 1-\frac{j}{n} \right)\bw_i^T\bSigma_1\left[\bSigma_{\by, \varepsilon_{i}}(j)+\bSigma_{\by, \varepsilon_{i}}^T(j)\right] \bSigma_0\be_i ,
\end{split}\nonumber
\end{equation}
and $\sup_{p}\sum_{j=1}^{\infty}|\bw_i^T\bSigma_1\bSigma_{\by, \varepsilon_{i}}(j) \bSigma_0\be_i| < \infty$. Calculating all the variance and covariance and summing up them, it follows from dominate convergence theorem that
\begin{equation}
\mathrm{Var}\left(\frac{S_{n,p}}{\sqrt{\ba^{T}\bU_i\ba}}\right) \to 1. \nonumber
\end{equation}

To prove the asymptotic normality of $S_{n,p}$, we employ the small-block and large-block arguments. We partition the set $\{1,2,\cdots,n\}$ into $2k_n+1$ subsets with large blocks of size $l_n$, small blocks of size $s_n$ and the last remaining set of size $n-k_n(l_n+s_n)$. Put
\[
l_n=[\sqrt{n}/\log{n}], \quad  s_n=[\sqrt{n} \log{n}]^x, \quad k_n=[n/(l_n+s_n)],
\]
where $ \frac{\gamma}{4+\gamma} \le x <1$. Hence
\[
l_n / \sqrt{n} \to 0, \quad s_n / l_n \to 0, \quad k_n=O(\sqrt{n} \log{n}).
\]
Note that $l_n / \sqrt{n} \to 0$ is important when we derive the Lindeberg condition of the truncated partial sum $T_{n,p}^L$ defined in (\ref{truncation}).

Since $\sum_{j=1}^{\infty}\alpha(j)^{\frac{\gamma}{4+\gamma}} < \infty$, we have $\alpha(s_n)=o(s_n^{-\frac{4+\gamma}{\gamma}})$. It then holds that
\[
k_n\alpha(s_n)=o(k_n / s_n^{\frac{4+\gamma}{\gamma}})=o(\sqrt{n} \log{n} / [\sqrt{n}\log{n}]^{x\frac{4+\gamma}{\gamma}})=o(1).
\]
Then we can partition $S_{n,p}$ in the following way
\begin{equation}
\begin{split}
S_{n,p}= &
a_1\frac{1}{\sqrt{n}}\sum_{j=1}^{k_n}\xi_j^{(1)}+a_2\frac{1}{\sqrt{n}}\sum_{j=1}^{k_n}\xi_j^{(2)}+a_3\frac{1}{\sqrt{n}}\sum_{j=1}^{k_n}\xi_j^{(3)}
\\ & +a_1\frac{1}{\sqrt{n}}\sum_{j=1}^{k_n}\eta_j^{(1)}+a_2\frac{1}{\sqrt{n}}\sum_{j=1}^{k_n}\eta_j^{(2)}+a_3\frac{1}{\sqrt{n}}\sum_{j=1}^{k_n}\eta_j^{(3)}
\\ & +a_1\frac{1}{\sqrt{n}}\zeta^{(1)}+a_2\frac{1}{\sqrt{n}}\zeta^{(2)}+a_3\frac{1}{\sqrt{n}}\zeta^{(3)}, \nonumber
\end{split}
\end{equation}
where
\[
\xi_j^{(1)}=\sum_{t=(j-1)(l_n+s_n)+1}^{jl_n+(j-1)s_n}\bw_i^T\bSigma_1\by_{t-1}\varepsilon_{i,t}, \quad \eta_j^{(1)}=\sum_{t=jl_n+(j-1)s_n+1}^{j(l_n+s_n)}\bw_i^T\bSigma_1\by_{t-1}\varepsilon_{i,t},
\]
\[
\xi_j^{(2)}=\sum_{t=(j-1)(l_n+s_n)+1}^{jl_n+(j-1)s_n}\be_i^T\bSigma_0\by_{t-1}\varepsilon_{i,t}, \quad \eta_j^{(2)}=\sum_{t=jl_n+(j-1)s_n+1}^{j(l_n+s_n)}\be_i^T\bSigma_0\by_{t-1}\varepsilon_{i,t},
\]
\[
\xi_j^{(3)}=\sum_{t=(j-1)(l_n+s_n)+1}^{jl_n+(j-1)s_n}\bw_i^T\bSigma_0\by_{t-1}\varepsilon_{i,t}, \quad \eta_j^{(3)}=\sum_{t=jl_n+(j-1)s_n+1}^{j(l_n+s_n)}\bw_i^T\bSigma_0\by_{t-1}\varepsilon_{i,t},
\]
\[
\zeta^{(1)}=\sum_{k_n(l_n+s_n)+1}^n\bw_i^T\bSigma_1\by_{t-1}\varepsilon_{i,t}, \quad \zeta^{(2)}=\sum_{k_n(l_n+s_n)+1}^n\be_i^T\bSigma_0\by_{t-1}\varepsilon_{i,t}, \quad \zeta^{(3)}=\sum_{k_n(l_n+s_n)+1}^n\bw_i^T\bSigma_0\by_{t-1}\varepsilon_{i,t}.
\]
Note that $\alpha(n)=o(n^{-\frac{(2+\gamma/2)2}{2(2+\gamma/2-2)}})$ and $k_ns_n/n \to 0$, $(l_n+s_n)/n \to 0$, by applying proposition 2.7 of Fan and Yao (2003), it holds that
\[
\frac{1}{\sqrt{n}}\sum_{j=1}^{k_n}\eta_j^{(l)}=o_p(1), \quad \text{and} \quad \frac{1}{\sqrt{n}}\zeta^{(l)}=o_p(1), \quad l=1,2,3.
\]
Therefore
\begin{equation}
S_{n,p}=a_1\frac{1}{\sqrt{n}}\sum_{j=1}^{k_n}\xi_j^{(1)}+a_2\frac{1}{\sqrt{n}}\sum_{j=1}^{k_n}\xi_j^{(2)}+a_3\frac{1}{\sqrt{n}}\sum_{j=1}^{k_n}\xi_j^{(3)}
+o_p(1) \equiv T_{n,p}+o_p(1). \nonumber
\end{equation}
We calculate the variance of $T_{n,p}$. Similar to (\ref{variance}), it holds that
\begin{equation}
\begin{split}
& \mathrm{Var}\left( a_1\frac{1}{\sqrt{n}}\sum_{j=1}^{k_n}\xi_j^{(1)} \right) =a_1^2\frac{k_n}{n}\var\left(\xi_1^{(1)}\right)\{1+o(1)\}=a_1^2\frac{k_n}{n}\mathrm{Var}\left(\sum_{t=1}^{l_n}\bw_i^T\bSigma_1\by_{t-1}\varepsilon_{i,t}
\right)\{1+o(1)\}
\\ = & a_1^2\frac{k_nl_n}{n}\left[ \bw_i^T\bSigma_1\bSigma_{\by, \varepsilon_{i}}(0) \bSigma_1^{T}\bw_i+\sum_{j=1}^{l_n-1}\left( 1-\frac{j}{l_n} \right)\bw_i^T\bSigma_1\left[\bSigma_{\by, \varepsilon_{i}}(j) + \bSigma_{\by, \varepsilon_{i}}^T(j)\right]\bSigma_1^{T}\bw_i \right]\{1+o(1)\}.
\end{split}\nonumber
\end{equation}
Calculating all the variance and covariance and summing up them, by dominated convergence theorem and $\frac{k_nl_n}{n} \to 1$, it holds that
\begin{equation}
\mathrm{Var}\left(\frac{T_{n,p}}{\sqrt{\ba^{T}\bU_i\ba}}\right) \to 1. \nonumber
\end{equation}
Now it suffices to prove the asymptotic normality of $T_{n,p}$. We partition $T_{n,p}$ into two parts via truncation. Specifically, we define
\[
\xi_j^{(1)L}=\sum_{t=(j-1)(l_n+s_n)+1}^{jl_n+(j-1)s_n}\bw_i^T\bSigma_1\by_{t-1}\varepsilon_{i,t}I_{\{|\bw_i^T\bSigma_1
\by_{t-1}\varepsilon_{i,t}| \le L\}},
\]
and
\[
\xi_j^{(1)R}=\sum_{t=(j-1)(l_n+s_n)+1}^{jl_n+(j-1)s_n}\bw_i^T\bSigma_1\by_{t-1}\varepsilon_{i,t}I_{\{|\bw_i^T\bSigma_1
\by_{t-1}\varepsilon_{i,t}| > L\}}.
\]
Similarly, we have $\xi_j^{(2)L}, \xi_j^{(2)R}$ and $\xi_j^{(3)L}, \xi_j^{(3)R}$. Then
\begin{equation} \label{truncation}
\begin{split}
T_{n,p}= & \left(a_1\frac{1}{\sqrt{n}}\sum_{j=1}^{k_n}\xi_j^{(1)L}+a_2\frac{1}{\sqrt{n}}\sum_{j=1}^{k_n}\xi_j^{(2)L}+
a_3\frac{1}{\sqrt{n}}\sum_{j=1}^{k_n}\xi_j^{(3)L}\right)
\\ & +\left(a_1\frac{1}{\sqrt{n}}\sum_{j=1}^{k_n}\xi_j^{(1)R}+a_2\frac{1}{\sqrt{n}}\sum_{j=1}^{k_n}\xi_j^{(2)R}+
a_3\frac{1}{\sqrt{n}}\sum_{j=1}^{k_n}\xi_j^{(3)R}\right)
\\ \equiv & T_{n,p}^L+T_{n,p}^R.
\end{split} \tag{16}
\end{equation}
Similar to computing the $\var(T_{n,p})$, it holds that
\begin{equation}
\begin{split}
& \var(T_{n,p}^L) =a_1^2\var\left( \frac{1}{\sqrt{n}}\sum_{j=1}^{k_n}\xi_j^{(1)L} \right)+ \Omega^L= a_1^2\frac{k_n}{n}\var\left(\xi_1^{(1)L}\right)\{1+o(1)\}+ \Omega^L
\\ = &  a_1^2\frac{k_n}{n}\var\left(\sum_{t=1}^{l_n}\bw_i^T\bSigma_1\by_{t-1}\varepsilon_{i,t}I_{\{|\bw_i^T\bSigma_1
\by_{t-1}\varepsilon_{i,t}| \le L\}}\right)\{1+o(1)\}+ \Omega^L
\\ = &  a_1^2 \frac{k_nl_n}{n}\Bigg[\var\left(\bw_i^T\bSigma_1\by_{t-1}\varepsilon_{i,t}I_{\{|\bw_i^T\bSigma_1
\by_{t-1}\varepsilon_{i,t}| \le L\}}\right)
\\ & +2\sum_{j=1}^{l_n-1}\left( 1-\frac{j}{l_n} \right) \cov\left(
\bw_i^T\bSigma_1\by_{t-1+j}\varepsilon_{i,t+j}I_{\{|\bw_i^T\bSigma_1
\by_{t-1+j}\varepsilon_{i,t+j}| \le L\}}, \bw_i^T\bSigma_1\by_{t-1}\varepsilon_{i,t}I_{\{|\bw_i^T\bSigma_1
\by_{t-1}\varepsilon_{i,t}| \le L\}}\right)\Bigg]
\\ & \{1+ o(1)\}+ \Omega^L,
\end{split} \nonumber
\end{equation}
where $\Omega^L$ is the sum of all the rest variance and covariance except $\var\left(a_1 \frac{1}{\sqrt{n}}\sum_{j=1}^{k_n}\xi_j^{(1)L} \right)$. Therefore
\[
\var\left(\frac{\var(T_{n,p}^L)}{\sigma_L^{2}}\right) \to 1,
\]
where we denote $\sigma_L^{2}$ as the asymptotic variance of $T_{n,p}^L$. Similarly, we have
\begin{equation}
\begin{split}
& \var(T_{n,p}^R)
\\ = & a_1^2 \frac{k_nl_n}{n}\Bigg[\var\left(\bw_i^T\bSigma_1\by_{t-1}\varepsilon_{i,t}I_{\{|\bw_i^T\bSigma_1
\by_{t-1}\varepsilon_{i,t}| > L\}}\right)
\\ & +2\sum_{j=1}^{l_n-1}\left( 1-\frac{j}{l_n} \right) \cov\left(
\bw_i^T\bSigma_1\by_{t-1+j}\varepsilon_{i,t+j}I_{\{|\bw_i^T\bSigma_1
\by_{t-1+j}\varepsilon_{i,t+j}| > L\}}, \bw_i^T\bSigma_1\by_{t-1}\varepsilon_{i,t}I_{\{|\bw_i^T\bSigma_1
\by_{t-1}\varepsilon_{i,t}| > L\}}\right)\Bigg]
\\ & \{1+ o(1)\}+ \Omega^R.
\end{split} \nonumber
\end{equation}
Define
\[
M_{n,p}=\left|\mathrm{E}\exp{\left(\frac{itT_{n,p}}{\sqrt{\ba^{T}\bU_i\ba}}\right)}-\exp{\left(-\frac{t^2}{2}\right)}\right|,
\]
where $i=\sqrt{-1}$ now. We bound $M_{n,p}$ as follows
\begin{equation}
\begin{split}
M_{n,p}  \le & \mathrm{E}\left|\exp{\left(\frac{itT_{n,p}^L}{\sqrt{\ba^{T}\bU_i\ba}}\right)}\left[\exp{\left(\frac{itT_{n,p}^R}{\sqrt{\ba^{T}
\bU_i\ba}}\right)}-1\right]\right|
\\  & + \left|\mathrm{E}\exp{\left(\frac{itT_{n,p}^L}{\sqrt{\ba^{T}\bU_i\ba}}\right)}-\prod\limits_{j=1}^{k_n}\mathrm{E}\exp{\left[\frac{it\left(a_1\frac{1}
{\sqrt{n}}\xi_j^{(1)L}+a_2\frac{1}{\sqrt{n}}\xi_j^{(2)L}+a_3\frac{1}{\sqrt{n}}\xi_j^{(3)L}\right)}{\sqrt{\ba^{T}\bU_i\ba}}
\right]}\right|
\\ & + \left|\prod\limits_{j=1}^{k_n}\mathrm{E}\exp{\left[\frac{it\left(a_1\frac{1}
{\sqrt{n}}\xi_j^{(1)L}+a_2\frac{1}{\sqrt{n}}\xi_j^{(2)L}+a_3\frac{1}{\sqrt{n}}\xi_j^{(3)L}\right)}{\sqrt{\ba^{T}\bU_i\ba}}
\right]}-\exp{\left(-\frac{t^2}{2}\frac{\sigma_L^2}{\ba^{T}\bU_i\ba}\right)}\right|
\\ & + \left|\exp{\left(-\frac{t^2}{2}\frac{\sigma_L^2}{\ba^{T}\bU_i\ba}\right)}-
\exp{\left(-\frac{t^2}{2}\right)}\right|.
\end{split} \nonumber
\end{equation}
Following the same arguments as part 2.7.7 of Fan and Yao (2003), for any $\epsilon>0$, it holds that $M_{n,p} < \epsilon$ as $n, p \to \infty$. Hence
\[
\sqrt{n} \times \ba^{T}\left(
\begin{array}{c}
\frac{1}{n}\sum_{t=1}^{n}\by_{t-1}^T(\bw_i^T\by_t) \frac{1}{n}\sum_{t=1}^{n}\varepsilon_{i, t}\by_{t-1}  \\
\frac{1}{n}\sum_{t=1}^{n}\by_{t-1}^Ty_{i,t-1} \frac{1}{n}\sum_{t=1}^{n}\varepsilon_{i, t}\by_{t-1} \\
\frac{1}{n}\sum_{t=1}^{n}\by_{t-1}^T(\bw_i^T\by_{t-1}) \frac{1}{n}\sum_{t=1}^{n}\varepsilon_{i, t}\by_{t-1}
\end{array}
\right)/\sqrt{\ba^{T}\bU_i\ba} \xrightarrow{d} N(0,1).
\]
Substituting $\ba$ by $(\bU_i^{-\frac{1}{2}})^T\ba$, it holds that
\[
\ba^{T}\left\{ \sqrt{n} \bU_i^{-\frac{1}{2}}  \left(
\begin{array}{c}
\frac{1}{n}\sum_{t=1}^{n}\by_{t-1}^T(\bw_i^T\by_t) \frac{1}{n}\sum_{t=1}^{n}\varepsilon_{i, t}\by_{t-1}  \\
\frac{1}{n}\sum_{t=1}^{n}\by_{t-1}^Ty_{i,t-1} \frac{1}{n}\sum_{t=1}^{n}\varepsilon_{i, t}\by_{t-1} \\
\frac{1}{n}\sum_{t=1}^{n}\by_{t-1}^T(\bw_i^T\by_{t-1}) \frac{1}{n}\sum_{t=1}^{n}\varepsilon_{i, t}\by_{t-1}
\end{array}
\right)
\right\}
\xrightarrow{d} \ba^{T} N(0,\bI_3),
\]
which leads to the fact that
\[
\sqrt{n} \bU_i^{-\frac{1}{2}}
\left(
\begin{array}{c}
\frac{1}{n}\sum_{t=1}^{n}\by_{t-1}^T(\bw_i^T\by_t) \frac{1}{n}\sum_{t=1}^{n}\varepsilon_{i, t}\by_{t-1}  \\
\frac{1}{n}\sum_{t=1}^{n}\by_{t-1}^Ty_{i,t-1} \frac{1}{n}\sum_{t=1}^{n}\varepsilon_{i, t}\by_{t-1} \\
\frac{1}{n}\sum_{t=1}^{n}\by_{t-1}^T(\bw_i^T\by_{t-1}) \frac{1}{n}\sum_{t=1}^{n}\varepsilon_{i, t}\by_{t-1}
\end{array}
\right)
\xrightarrow{d} N(0,\bI_3).
\]

To prove (2), let us look at the $(1,1)$-th element of $\hat \bX_i^T\hat \bX_i$. We have
\begin{equation} \label{decom_decom}
\begin{split}
& \frac{1}{n}\sum_{t=1}^{n}\by_{t-1}^T(\bw_i^T\by_t)\frac{1}{n}\sum_{t=1}^{n}\by_{t-1}(\bw_i^T\by_t)
\\ = & \left(\frac{1}{n}\sum_{t=1}^{n}\by_{t-1}^T(\bw_i^T\by_t)-\bw_i^{T}\bSigma_1\right)\left(\frac{1}{n}\sum_{t=1}^{n}\by_{t-1}
(\bw_i^T\by_t)-\bSigma_1^{T}\bw_i\right)
\\ & + 2\bw_i^{T}\bSigma_1\left(\frac{1}{n}\sum_{t=1}^{n}\by_{t-1}
(\bw_i^T\by_t)-\bSigma_1^{T}\bw_i\right)+\bw_i^{T}\bSigma_1\bSigma_1^{T}\bw_i.
\end{split} \tag{17}
\end{equation}
Using the same arguments as (\ref{mixing_bound}), the first term is $O_p({p \over n})$ and the second term is $O_p({1 \over \sqrt{n}})$. Hence given $p=o(n)$, it holds that
\[
\frac{\frac{1}{n}\sum_{t=1}^{n}\by_{t-1}^T(\bw_i^T\by_t)\frac{1}{n}
\sum_{t=1}^{n}\by_{t-1}(\bw_i^T\by_t)}{\bw_i^{T}\bSigma_1\bSigma_1^{T}\bw_i} \to 1.
\]
Applying the same arguments to the other elements of $\hat \bX_i^T\hat \bX_i$, it holds that
\[
\bV_i (\hat \bX_i^T\hat \bX_i)^{-1} \xrightarrow{P} \bI_3.
\]

To prove (ii) in Theorem 2,
the required asymptotic result follows from (\ref{error_decom}) and (\ref{decom_decom})
immediately when $p=o(n)$ and $\sqrt{n}=O(p)$.
The proof is completed.
\qqed

\noindent
{\bf Proof of Corollary 1.}
By Theorem 2, it holds that
\begin{equation}
\left\|
\left(
\begin{array}{c}
\hat\lambda_{0i} \\
\hat\lambda_{1i} \\
\hat\lambda_{2i}
\end{array}
\right)
-
\left(
\begin{array}{c}
\lambda_{0i} \\
\lambda_{1i} \\
\lambda_{2i}
\end{array}
\right)\right\|_1 =
\begin{cases}
O_p(\frac{1}{\sqrt{n}}) \quad & {\rm if} \;\; \frac{p}{\sqrt{n}}=O(1), \\
O_p(\frac{p}{n}) & {\rm if} \;\; \frac{p}{\sqrt{n}} \to \infty \;\; {\rm and} \;\; \frac{p}{n}=o(1). \nonumber
\end{cases}
\end{equation}
for all $i$. The required asymptotic result follows from the above result directly.
\qqed

\noindent
{\bf Proof of Theorem 4.}
Let us look at term $E_1$ and $E_2$ in (\ref{error_decom}) first under the new condition (A5). Similar to the proof of (\ref{mixing_bound}), it holds that
\[
E_1=O_p(\frac{p s_1^{3/4}(p)}{n}), \quad E_2=O_p(\frac{s_0^{1/4}(p)}{\sqrt{n}}).
\]
Hence
\begin{equation}
\begin{split}
& \frac{1}{n}\sum_{t=1}^n \by_{t-1}^{T}(\bw_i^T\by_t)\frac{1}{n}\sum_{t=1}^n \varepsilon_{i,t}\by_{t-1}=O_p(\frac{p s_1^{3/4}(p)}{n}+\frac{s_0^{1/4}(p)}{\sqrt{n}}).
\end{split} \nonumber
\end{equation}
Similarly, we have
\begin{equation}
\begin{split}
& \frac{1}{n}\sum_{t=1}^{n}\by_{t-1}^Ty_{i,t-1}  \frac{1}{n}\sum_{t=1}^{n}\varepsilon_{i, t}\by_{t-1}=O_p(\frac{p s_1^{3/4}(p)}{n}+\frac{s_0^{1/4}(p)}{\sqrt{n}}),
\end{split} \nonumber
\end{equation}
\begin{equation}
\begin{split}
& \frac{1}{n}\sum_{t=1}^{n}\by_{t-1}^T(\bw_i^T\by_{t-1})  \frac{1}{n}\sum_{t=1}^{n}\varepsilon_{i, t}\by_{t-1}=O_p(\frac{p s_1^{3/4}(p)}{n}+\frac{s_0^{1/4}(p)}{\sqrt{n}}).
\end{split} \nonumber
\end{equation}

For the first diagonal element of $\hat \bX_i^T\hat \bX_i$, it follows from considering the three terms in (\ref{decom_decom}) separately that
\[
\frac{1}{n}\sum_{t=1}^{n}\by_{t-1}^T(\bw_i^T\by_t)\frac{1}{n}\sum_{t=1}^{n}\by_{t-1}(\bw_i^T\by_t)=O_p(\frac{p s_1(p)}{n}+\frac{s_0^{1/4}(p)s_1^{1/4}(p)}{\sqrt{n}})+\bw_i^{T}\bSigma_1\bSigma_1^{T}\bw_i.
\]
Similarly,
\[
\frac{1}{n}\sum_{t=1}^{n}\by_{t-1}^Ty_{i,t-1}\frac{1}{n}\sum_{t=1}^{n}\by_{t-1}y_{i,t-1}=O_p(\frac{p s_1(p)}{n}+\frac{s_0^{1/4}(p)s_1^{1/4}(p)}{\sqrt{n}})+\be_i^{T}\bSigma_0\bSigma_0^{T}\be_i,
\]
\[
\frac{1}{n}\sum_{t=1}^{n}\by_{t-1}^T(\bw_i^T\by_{t-1}) \frac{1}{n}\sum_{t=1}^{n}\by_{t-1}(\bw_i^T\by_{t-1})=O_p(\frac{p s_1(p)}{n}+\frac{s_0^{1/4}(p)s_1^{1/4}(p)}{\sqrt{n}})+\bw_i^{T}\bSigma_0\bSigma_0^{T}\bw_i.
\]

Given $\frac{ps_1(p)}{s_2(p)}=o(n)$ and $\frac{s_0^{1/2}(p)}{ps_1^{1/2}(p)s_2(p)}=O(1)$, we have
\[
\frac{p s_1(p)}{n}=o(s_2(p)), \quad \frac{s_0^{1/4}(p)s_1^{1/4}(p)}{\sqrt{n}}=o(s_2(p)).
\]
Divide both the numerator and denominator of estimator (\ref{estimator}) by $s_2(p)$, it holds that
\begin{equation}
\left\|
\left(
\begin{array}{c}
\hat\lambda_{0i} \\
\hat\lambda_{1i} \\
\hat\lambda_{2i}
\end{array}
\right)
-
\left(
\begin{array}{c}
\lambda_{0i} \\
\lambda_{1i} \\
\lambda_{2i}
\end{array}
\right)\right\|_2 =
O_p\Big(\frac{p s_1^{3/4}(p)}{ns_2(p)}+\frac{s_0^{1/4}(p)}{\sqrt{n}s_2(p)}\Big).
\nonumber
\end{equation}
The required result then follows directly.
\qqed

\begin{lemma} \label{lemma1}
Under conditions A1 and B1 -- B3, condition A2 holds with $\gamma=4$.
\end{lemma}

\noindent\textbf{Proof.}
It is apparent that part (a) of A2 is satisfied under A1 and B1 -- B3. $\by_t$ is strictly stationary because $\varepsilon_{i, t}$ are $i.i.d$ across $i$ and $t$ and condition B3. Since the density function of $\varepsilon_{i, t}$ exists, $\alpha(n)$ decays exponentially fast, see Pham and Tran (1985). Therefore $
\sum_{j=1}^{\infty}\alpha(j)^{\frac{\gamma}{4+\gamma}} < \infty$.
Now we prove A2(c) when $\gamma=4$.

We present a more general result first: for any $p \times 1$ vector $\ba$ satisfying $\sup_{p}\|\ba\|_1 < \infty$, it holds that
\[
\sup_{p} \mathrm{E}\left|\ba^T\by_t\right|^{8} < \infty.
\]
Note that
\begin{equation}
\by_t=\sum_{h=0}^{\infty}\bA^h\bS^{-1}(\blambda_0)\bve_{t-h} \equiv \sum_{h=0}^{\infty}\bB_h\bve_{t-h}. \nonumber
\end{equation}
Then
\begin{equation} \label{ineq1}
\begin{split}
& \mathrm{E}\left|\ba^T\by_t\right|^{8}=\mathrm{E}\left|\sum_{h=0}^{\infty}\ba^T\bB_h\bve_{t-h}\right|^{8} \equiv \mathrm{E}\left|\sum_{h=0}^{\infty}\bb_h^T\bve_{t-h}\right|^{8}
\\ = & \mathrm{E}\left|\sum_{h_1, h_2, h_3, h_4, h_5, h_6, h_7, h_8=0}^{\infty}(\bve_{t-h_1}^T\bb_{h_1}\bb_{h_2}^T\bve_{t-h_2})(\bve_{t-h_3}^T\bb_{h_3}\bb_{h_4}^T\bve_{t-h_4})(\bve_{t-h_5}^T\bb_{h_5}
\bb_{h_6}^T\bve_{t-h_6})(\bve_{t-h_7}^T\bb_{h_7}\bb_{h_8}^T\bve_{t-h_8})\right|
\\ = & \mathrm{E} \Bigg| \sum_{h_1, h_2, h_3, h_4, h_5, h_6, h_7, h_8=0}^{\infty} \Big(\sum_{i_1, j_1=1}^p [\bb_{h_1}\bb_{h_2}^T]_{i_1 j_1} \varepsilon_{i_1, t-h_1} \varepsilon_{j_1, t-h_2}\Big)\Big(\sum_{i_2, j_2=1}^p [\bb_{h_3}\bb_{h_4}^T]_{i_2 j_2} \varepsilon_{i_2, t-h_3} \varepsilon_{j_2, t-h_4}\Big)
\\ & \times \Big(\sum_{i_3, j_3=1}^p [\bb_{h_5}\bb_{h_6}^T]_{i_3 j_3} \varepsilon_{i_3, t-h_5} \varepsilon_{j_3, t-h_6}\Big)\Big(\sum_{i_4, j_4=1}^p [\bb_{h_7}\bb_{h_8}^T]_{i_4 j_4} \varepsilon_{i_4, t-h_7} \varepsilon_{j_4, t-h_8}\Big) \Bigg|
\\ = & \mathrm{E} \Bigg| \sum_{h_1, h_2, h_3, h_4, h_5, h_6, h_7, h_8=0}^{\infty} \quad \sum_{i_1, j_1, i_2, j_2, i_3, j_3, i_4, j_4=1}^p [\bb_{h_1}\bb_{h_2}^T]_{i_1 j_1}[\bb_{h_3}\bb_{h_4}^T]_{i_2 j_2} [\bb_{h_5}\bb_{h_6}^T]_{i_3 j_3}[\bb_{h_7}\bb_{h_8}^T]_{i_4 j_4}
\\ & \times \varepsilon_{i_1, t-h_1} \varepsilon_{j_1, t-h_2} \varepsilon_{i_2, t-h_3} \varepsilon_{j_2, t-h_4}
\varepsilon_{i_3, t-h_5} \varepsilon_{j_3, t-h_6}  \varepsilon_{i_4, t-h_7} \varepsilon_{j_4, t-h_8} \Bigg|
\\ \le &  \sum_{h_1, h_2, h_3, h_4, h_5, h_6, h_7, h_8=0}^{\infty} \quad \sum_{i_1, j_1, i_2, j_2, i_3, j_3, i_4, j_4=1}^p
\Big| [\bb_{h_1}\bb_{h_2}^T]_{i_1 j_1}[\bb_{h_3}\bb_{h_4}^T]_{i_2 j_2} [\bb_{h_5}\bb_{h_6}^T]_{i_3 j_3}[\bb_{h_7}\bb_{h_8}^T]_{i_4 j_4} \Big|
\\ & \times \mathrm{E} |\varepsilon_{i_1, t-h_1} \varepsilon_{j_1, t-h_2} \varepsilon_{i_2, t-h_3} \varepsilon_{j_2, t-h_4}
\varepsilon_{i_3, t-h_5} \varepsilon_{j_3, t-h_6}  \varepsilon_{i_4, t-h_7} \varepsilon_{j_4, t-h_8}|
\\ \le & C\sum_{h_1, h_2, h_3, h_4, h_5, h_6, h_7, h_8=0}^{\infty} \quad \sum_{i_1, j_1, i_2, j_2, i_3, j_3, i_4, j_4=1}^p
|\bb_{h_1}\bb_{h_2}^T|_{i_1 j_1}|\bb_{h_3}\bb_{h_4}^T|_{i_2 j_2} |\bb_{h_5}\bb_{h_6}^T|_{i_3 j_3}|\bb_{h_7}\bb_{h_8}^T|_{i_4 j_4}
\\ = & C \Big[\sum_{h=0}^{\infty}\sum_{g=0}^{\infty}\sum_{i=1}^p\sum_{j=1}^p|\bb_{h}\bb_{g}^T|_{i j}\Big]^4.
\end{split} \tag{18}
\end{equation}
And
\begin{equation} \label{ineq2}
\begin{split}
& \sum_{h=0}^{\infty}\sum_{g=0}^{\infty}\sum_{i=1}^p\sum_{j=1}^p|\bb_{h}\bb_{g}^T|_{i j} \le \sum_{h=0}^{\infty}\sum_{g=0}^{\infty}\sum_{i=1}^p\sum_{j=1}^p(|\bb_{h}||\bb_{g}^T|)_{i j}
=\sum_{i=1}^p\sum_{j=1}^p \Big(\sum_{h=0}^{\infty}\sum_{g=0}^{\infty}|\bb_{h}||\bb_{g}^T|\Big)_{ij}
\\ = &  \sum_{i=1}^p\sum_{j=1}^p \Big(\sum_{h=0}^{\infty}|\bb_{h}|\sum_{g=0}^{\infty}|\bb_{g}^T|\Big)_{ij}
=\sum_{i=1}^p\sum_{j=1}^p \Big(\sum_{h=0}^{\infty}|\bb_{h}|\Big)_i\Big(\sum_{g=0}^{\infty}|\bb_{g}|\Big)_j
\\ = & \sum_{i=1}^p\Big(\sum_{h=0}^{\infty}|\bb_{h}|\Big)_i \sum_{j=1}^p \Big(\sum_{g=0}^{\infty}|\bb_{g}|\Big)_j,
\end{split} \tag{19}
\end{equation}
where $\Big(\sum_{h=0}^{\infty}|\bb_{h}|\Big)_i$ is the $i$-th element of the column vector $\sum_{h=0}^{\infty}|\bb_{h}|$.

Since $\left(\sum_{h=0}^{\infty}\left|\bB_h\right|\right)_{ij}=\sum_{h=0}^{\infty}\left(\left|\bA^h
\bS^{-1}(\blambda_0)\right|\right)_{ij}\le \left(\sum_{h=0}^{\infty}\left|\bA^h\right|\left|
\bS^{-1}(\blambda_0)\right|\right)_{ij}$ where the row and column sums of $\sum_{h=0}^{\infty}\left|\bA^h\right|\left| \bS^{-1}(\blambda_0)\right|$ are bounded uniformly in $p$, it holds that the row and column sums of $\sum_{h=0}^{\infty}\left|\bB_{h}\right|$ are bounded uniformly in $p$. Note that
\[
\Big(\sum_{h=0}^{\infty}|\bb_{h}|\Big)_i=\Big(\sum_{h=0}^{\infty}|\bB_{h}^T\ba|\Big)_i\le \Big(\sum_{h=0}^{\infty}|\bB_{h}^T||\ba|\Big)_i,
\]
where the row and column sums of $\sum_{h=0}^{\infty}\left|\bB_{h}^T\right|$ and $|\ba|$ are bounded uniformly in $p$. Hence the row and column sums of $\sum_{h=0}^{\infty}|\bB_{h}^T||\ba|$ are bounded uniformly in $p$. It follows from (\ref{ineq1}) and (\ref{ineq2}) that
\[
\sup_{p}\mathrm{E}\left|\ba^T\by_t\right|^{8} \le C \Big[\sum_{i=1}^p\Big(\sum_{h=0}^{\infty}|\bb_{h}|\Big)_i \sum_{j=1}^p \Big(\sum_{g=0}^{\infty}|\bb_{g}|\Big)_j\Big]^4=O(1).
\]
It is easy to prove that
\[
\sup_{p} \|\bSigma_0\bw_i\|_1 <\infty, \quad \sup_{p} \|\bSigma_1^T\bw_i\|_1 <\infty, \quad \sup_{p} \|\bSigma_0\be_i\|_1 <\infty.
\]
Thus $\sup_{p}\|\bw_i\bSigma_0\by_t\|_1 < \infty$ and etc.

The row and column sums of $\bSigma_0$ and $\bSigma_1$ are bounded uniformly in $p$. Then
\[
\sup_{p}\bw_i^{T}\bSigma_1\bSigma_1^{T}\bw_i=O(1).
\]
Similarly, we can prove the other diagonal elements of $\bV_i$ and $\bU_i$ are bounded uniformly in $p$. \\
The proof is completed.
\qqed

\section*{Acknowledgements}
Maria Lucia Parrella was partially supported by
 the Italian Ministry of Education, University
and Research (MIUR), PRIN Research Project 2010-2011 - prot. 2010J3LZEN,
``Forecasting economic and financial time series: understanding the
complexity and modelling structural change''. Qiwei Yao was partially supported
by an EPSRC research grant.

\end{document}